\documentclass[11pt]{iopart}
\usepackage[utf8]{inputenc}
\usepackage{url}
\usepackage{graphicx}
\usepackage{amssymb}
\usepackage{amsfonts,amssymb}
\usepackage{graphicx}
\usepackage{lmodern}
\usepackage{multicol}
\usepackage{hyperref} 
\usepackage[mathscr]{euscript}
\usepackage{braket}
\usepackage[margin=1in]{geometry}
\usepackage{lscape}
\usepackage{caption}
\usepackage{subcaption}
\usepackage[toc]{appendix}
\usepackage{multirow}
\usepackage[usenames, dvipsnames]{color}
\usepackage{xspace}
\newcommand{\gy}{\textsc{Gyoto}\xspace}
\newcommand{\dd}{\mathrm{d}}

\makeatletter
\newcounter{parentsubcaption}
\newenvironment{subsubcaption}
 {\refstepcounter{sub\@captype}%
  \protected@edef\theparentsubcaption{\@nameuse{thesub\@captype}}%
  \setcounter{parentsubcaption}{\value{sub\@captype}}%
  \setcounter{sub\@captype}{0}%
  \@namedef{thesub\@captype}{\theparentsubcaption--\arabic{sub\@captype}}%
  \ignorespaces
}{%
  \setcounter{sub\@captype}{\value{parentsubcaption}}%
  \ignorespacesafterend
}
\makeatother

\begin{document}
\title{Imaging a non-singular rotating black hole at the center of the Galaxy}
\author{F. Lamy$^1$, E. Gourgoulhon$^2$, T. Paumard$^{3}$, F. H. Vincent$^{3}$}
\address{$^1$ AstroParticule et Cosmologie, Universit\'e Paris Diderot, CNRS, CEA, Observatoire de Paris, Sorbonne Paris Cit\'e.
B\^atiment Condorcet, 10 rue Alice Domon et L\'eonie Duquet, F-75205 Paris Cedex 13, France}
\address{$^2$
LUTH, Observatoire de Paris, Universit\'e PSL, CNRS, Universit\'e Paris Diderot, Sorbonne Paris Cit\'e,
5 place Jules Janssen, 92190 Meudon, France}
\address{$^3$ LESIA, Observatoire de Paris, Université PSL, CNRS, Sorbonne Université, Univ. Paris Diderot, Sorbonne Paris Cité, 5 place Jules Janssen, 92195 Meudon, France}
\ead{frederic.lamy@apc.in2p3.fr, eric.gourgoulhon@obspm.fr, thibaut.paumard@obspm.fr, frederic.vincent@obspm.fr}

\def\aligned{\vcenter\bgroup\let\\\cr
\halign\bgroup&\hfil${}##{}$&${}##{}$\hfil\cr}
\def\endaligned{\crcr\egroup\egroup}

\begin{abstract}
We show that the rotating generalization of Hayward's non-singular black hole previously studied in the literature is geodesically incomplete, and that its straightforward extension leads to a singular spacetime. We present another extension, which is devoid of any curvature singularity. The obtained metric depends on three parameters and, depending on their values,
yields an event horizon or not. These two regimes, named respectively \emph{regular rotating Hayward black hole} and \emph{naked rotating wormhole}, are studied both numerically and analytically. In preparation for the upcoming results of the Event Horizon Telescope, the images of an accretion torus around Sgr~A*, the supermassive object at the center of the Galaxy, are computed. These images contain, even in the absence of a horizon, a central faint region which bears a resemblance to the shadow of Kerr black holes and emphasizes the difficulty of claiming the existence of an event horizon from the analysis of strong-field images. The frequencies of the co- and contra-rotating orbits at the innermost stable circular orbit (ISCO) in this geometry are also computed, in the hope that quasi-periodic oscillations may permit to compare this model with Kerr's black hole on observational grounds.

\end{abstract}

\submitto{\CQG}



\section{Introduction} \label{s:intro}

With the advent of the Event Horizon Telescope \cite{Doeleman:2017}
and the VLTI/GRAVITY instrument \cite{Abuter_al:2017}, black hole physics is
entering a new era, where observational tests of the celebrated no-hair theorem
of general relativity (see e.g. \cite{CardosoG:2016}) are becoming
feasible. This theorem states that the unique solution for a steady
isolated black hole in four-dimensional vacuum general relativity is the Kerr-Newman
black hole, which depends on only three parameters: the mass $M$,
the reduced angular momentum $a=J/M$ and the electric charge $Q$
(see \cite{IonescuK:2015} for a precise mathematical
statement). In the (astrophysically relevant)
electrically neutral case ($Q=0$), the solution reduces to the
Kerr black hole.
To prepare the observational tests, it is
primordial to compute observables from theoretically plausible alternatives to the
Kerr black hole (see \cite{Berti_al:2015} for a review).

Among the numerous alternatives, a large class is
constituted by the so-called \emph{non-singular black holes}, also named
\emph{regular black holes}, namely
asymptotically flat spacetimes with a black hole region (and hence an event horizon) but without any curvature singularity. This class circumvents the no-hair theorem because the
metrics are not vacuum solutions of Einstein equation.
The first non-singular black hole spacetime has been proposed by Bardeen in 1968 \cite{Bardeen:1968}.
It has been mentioned by Hawking and Ellis (p.~265 of Ref.~\cite{HawkingE:1973}),
while discussing the famous Penrose singularity theorem \cite{Penrose:1965}, since it provides an instructive
counter-example: all geodesics of Bardeen's spacetime are regular, despite it contains trapped surfaces and obeys the weak energy condition. Actually, this spacetime violates the third hypothesis of Penrose theorem in its original version \cite{Penrose:1965}: the existence of a Cauchy surface. It was shown later
by Ay\'on-Beato and Garc\'\i{}a \cite{AyonG:2000}
that the Bardeen metric is a solution of Einstein's equations
with the energy-momentum tensor arising from
a magnetic monopole in some nonlinear electrodynamics theory, thereby giving some
physical content to the model. Another famous regular black hole metric has
been proposed by Hayward in 2006 \cite{Hayward:2006}; it fulfills the weak
energy condition as well  and was shown recently to be
a solution of Einstein's equations corresponding to a magnetic monopole,
as the Bardeen black hole,
but in another nonlinear electrodynamics theory \cite{Fan:2017}.
Both Bardeen and Hayward black holes are actually part of a larger class of
solutions of Einstein's equations coupled to nonlinear electrodynamics
found recently by Fan and Wang \cite{Fan&Wang:2016} (see also \cite{Bronnikov:2017,Bronnikov:2001}).

The Bardeen and Hayward black holes, and more generally all solutions of the
Fan-Wang class, are spherically symmetric and static outside the event
horizon. Now, on astrophysical grounds, it sounds more relevant to
consider rotating black holes and even rapidly rotating ones (see e.g. the spin values
in Tables~I and II of
Ref.~\cite{Bambi:2017}).
The metrics of Bardeen and Hayward
have been generalized to rotating axisymmetric metrics
by Bambi and Modesto \cite{Bambi&Modesto:2013} via the Newman-Janis algorithm.
More generally, all metrics of the Fan-Wang class have been
recently extended
to rotating ones by Toshmatov et al. \cite{Toshmatov_et_al:2017,Toshmatov_et_al:2018}.
However, contrary to the nonrotating ones, the rotating metrics are only approximate solutions
describing a magnetic monopole in some nonlinear electrodynamics theory \cite{RodriguesJ:2017,Toshmatov_et_al:2018}. Another shortcoming of these spacetimes is being geodesically incomplete. Moreover, as we show below, their
straightforward extension leads to \emph{singular} black
holes, i.e. to spacetimes with a curvature singularity.

In this article, we apply a prescription devised by Torres \cite{Torres:2017}
to obtain a spacetime extended to negative values of the ``radial'' coordinate $r$
and representing a rotating non-singular
black hole that reduces to the Hayward solution in the nonrotating limit.
Moreover, we consider values in the parameter space for which the solution
is trully a black hole one, that we call \emph{regular rotating Hayward black hole}, but also those for which the solution
has no event horizon. In the Kerr case, this would correspond to
$a > M$ and would yield a naked singularity. In our case, the regularity
of spacetime is still preserved and we obtain instead a rotating
traversable wormhole configuration,
which we call a \emph{naked rotating wormhole}, to distinguish it from
other rotating wormholes introduced in the literature \cite{Teo:1998,ChewKK:2016,AbdujabbarovJAS:2016}.

The paper is organized as follows. In Sec. \ref{section 2}, we start by giving a review of Hayward's black hole and its generalization to nonzero rotation as introduced by Bambi and Modesto \cite{Bambi&Modesto:2013}. We show that this model still possesses a curvature singularity and apply Torres' prescription to obtain a regular rotating Hayward metric, whose features are investigated in detail. Sec. \ref{section 3} is devoted to the numerical study of this model, with or without horizons, using the ray-tracing code \gy~\cite{Vincent_et_al:2011}. Finally in Sec. \ref{section 4} we investigate analytically the existence of circular orbits of massive particles as well as the propagation of photons in this geometry. We give the explicit expressions for the specific energy and angular momentum of a particle on a co- or contra-rotating orbit, which differ from the expressions of Ref.~\cite{Toshmatov_et_al:2017}, and compute the frequencies of such particles at the innermost stable circular orbit (ISCO).

\section{Metric and motivations} \label{section 2}
\subsection{Spinning up Hayward's black hole}
\subsubsection{Hayward's original metric}
\leavevmode \par

A standard example of a non-singular static black hole of mass $m$
is provided by Hayward's metric \cite{Hayward:2006}:
\begin{equation} \label{e:Hayward_metric}
\dd s^2=- \left( 1-\frac{2M(r)}{r} \right) \dd t^2
 + \left( 1-\frac{2M(r)}{r} \right) ^{-1} \dd r^2
 +r^2 \,\dd\theta^2+r^2\sin^2\theta \, \dd\phi^2,
\end{equation}
with
\begin{equation} \label{e:def_F_M}
M(r):= m \frac{r^3}{r^3+2mb^2} ,
\end{equation}
where $m$ and $b$ are two constants having the dimension of a length (or a mass
in the geometrized units used here).
The metric (\ref{e:Hayward_metric}) reduces to Schwarzschild's metric of mass $m$
in the limit $r \rightarrow +\infty$, where the effective mass $M$ tends to $m$. Furthermore,
if $b\not=0$, we have $2M(r)/r \sim r^2/b^2$ for $r\rightarrow 0$, so that Hayward's metric
behaves as a de Sitter metric with cosmological constant $\Lambda=3/b^2$ around
$r=0$, thereby avoiding any singularity.

Hayward \cite{Hayward:2006} interpreted the parameter $b$ as a cut-off of the order of the Planck length, i.e. a length at which general relativity is no longer valid. An alternative interpretation
of $b$, allowing for macroscopic values, has been provided by Fan and Wang \cite{Fan&Wang:2016} (see also \cite{Bronnikov:2017}).
These authors have shown that the metric (\ref{e:Hayward_metric})-(\ref{e:def_F_M}) can be obtained as a solution
of Einstein's equations sourced by the energy-momentum tensor of a magnetic monopole
within some nonlinear electrodynamics. The parameter $b$ is then related to the
amplitude of the total magnetic charge $Q_{\rm mag}$ by
\begin{equation}
Q_{\rm mag} = \left( \frac{b m^4}{2} \right) ^{1/3} ,
\end{equation}
while the nonlinear electrodynamics theory is defined by the Lagrangian density
\begin{equation}
L = L(\mathcal{F}) := \frac{6}{b^2} \frac{(2b^2 \mathcal{F})^{3/2}}{(1+ (2b^2 \mathcal{F})^{3/4})^2},
\end{equation}
$\mathcal{F}$ being the invariant $\mathcal{F}:= F_{\mu\nu} F^{\mu\nu}$ of the electromagnetic
field. Note that standard (Maxwell) electrodynamics corresponds to $L(\mathcal{F}) = \mathcal{F}$.

\subsubsection{A first attempt to generalize Hayward's metric to nonzero rotation}
\leavevmode \par

By means of the Newman-Janis algorithm \cite{Newman&Janis:1965},
Bambi and Modesto \cite{Bambi&Modesto:2013} (see also \cite{TorresFayos:2017})
have obtained some rotating generalization of Hayward's metric as
\begin{equation} \label{metricKerrmodified}
\begin{aligned}
\dd s^2=&-\left(1-\frac{2rM(r)}{\Sigma} \right)\dd t^2-\frac{4arM(r)\sin^2 \theta}{\Sigma}\, \dd t \, \dd\phi + \frac{\Sigma}{\Delta} \, \dd r^2+\Sigma \, \dd\theta^2 \\
&+\sin^2 \theta \left(r^2+a^2+\frac{2a^2rM(r)\sin^2\theta}{\Sigma} \right) \dd \phi^2,
\end{aligned}
\end{equation}
where
\begin{equation}
\label{e:M_r_BambiModesto}
\begin{aligned}
 & \Sigma :=r^2+a^2\cos^2 \theta, \quad
 \Delta :=r^2-2M(r)r+a^2, \\
 & M(r) :=m \displaystyle \frac{r^3}{r^3+2mb^2}.
\end{aligned}
\end{equation}
In addition to the total mass $m$ and the characteristic length $b$,
the new parameter with respect to Hayward's metric (\ref{e:Hayward_metric})-(\ref{e:def_F_M}) is the spin parameter $a$, such that the total angular momentum
is $J = am$.
Note that the function $M(r)$ is identical to that defined by Eq.~(\ref{e:def_F_M})
and that, except for the dependency of $M$ with respect to $r$, the line element
(\ref{metricKerrmodified}) is identical to that of the Kerr metric expressed
in Boyer-Lindquist coordinates.

As claimed in Ref.~\cite{Bambi&Modesto:2013}, there is no singularity at $r=0$
as long as $b \neq 0$ (see \cite{TorresFayos:2017} for a rigorous proof).
However, as in the Kerr case, if the above
metric is limited to $r\geq 0$, it yields a spacetime that is not geodesically complete:
some timelike and null geodesics stop at $r=0$ for a finite value of their affine
parameter, while (i) there is no curvature singularity there and (ii) $r=0$ is not
a coordinate singularity as in Minkowski's spacetime. The last point can be seen
by considering the value of the metric (\ref{metricKerrmodified})-(\ref{e:M_r_BambiModesto}) at $r=0$:
\begin{equation}
\left. \dd s^2\right| _{r=0} = - \dd t^2 + \cos^2\theta \, \dd r^2
    + a^2\cos^2 \theta \, \dd\theta^2 +a^2 \sin^2 \theta \, \dd\phi^2.
\end{equation}
If $a\not=0$, this defines a regular (i.e. nondegenerate) metric, except for $\theta=\pi/2$,
the vanishing of $\sin^2\theta$ at $\theta=0$ or $\pi$ reflecting only the
standard coordinate singularity of spherical
coordinates on the rotation axis. The regularity of the metric at $r=0$ and the unphysical ending of geodesics
there leads one to extend the spacetime to negative values of $r$.
In other words, we consider
\begin{equation}
\mathscr{M} = \mathbb{R}^2\times\mathbb{S}^2
\end{equation}
as the spacetime manifold, with $(t,r)$ spanning $\mathbb{R}^2$ and
$(\theta,\phi)$ spanning the 2-sphere $\mathbb{S}^2$.

\begin{figure}[!h]
\begin{center}
\includegraphics[width=0.49\textwidth]{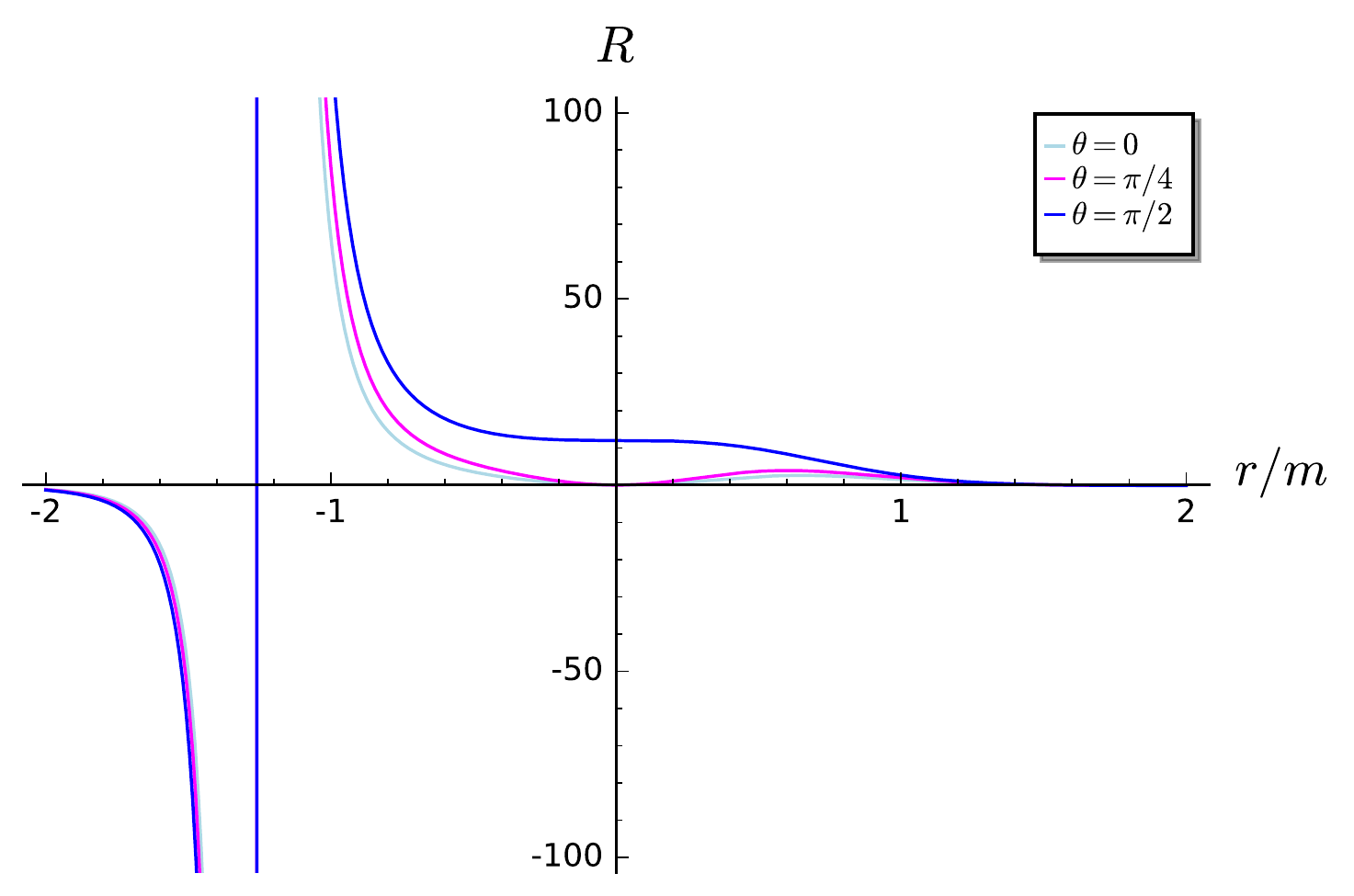}
\includegraphics[width=0.49\textwidth]{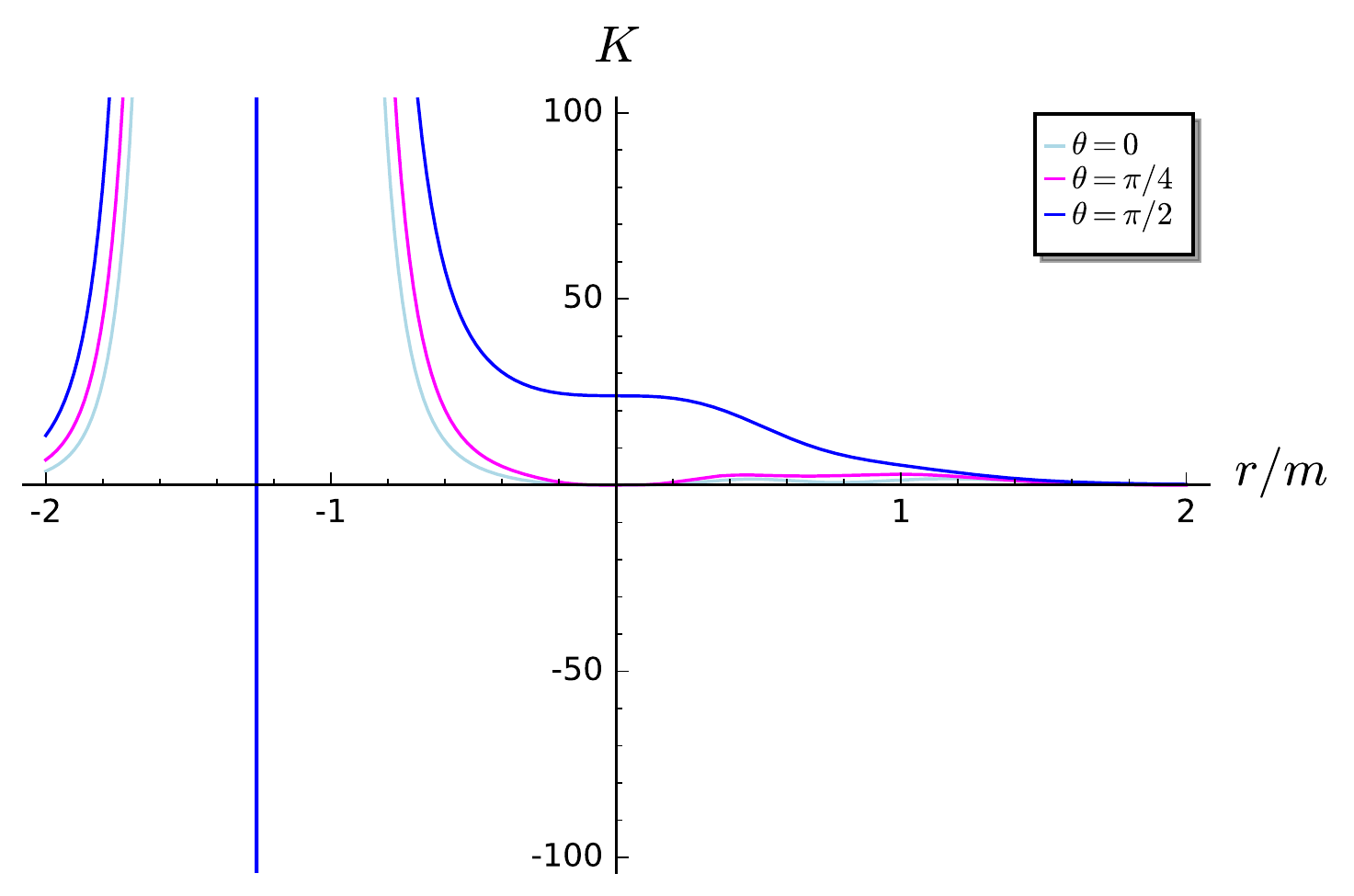}
\end{center}
\caption[]{\label{R&K_plots_H} \footnotesize
Ricci scalar (left) (in units of $m^{-2}$)
and Kretschmann scalar (right) (in units of $m^{-4}$) as functions of $r$
for the extension to $r<0$ of Bambi and Modesto \cite{Bambi&Modesto:2013}'s
rotating version of Hayward's metric with $a/m=0.9$ and $b/m=1$. Note that
both scalars are diverging at $r/m=-\sqrt[3]{2}\approx -1.26$.}
\end{figure}

Now, $\mathscr{M}$ endowed with the metric (\ref{metricKerrmodified})-(\ref{e:M_r_BambiModesto}) suffers from some curvature singularity, albeit not
at $r=0$.
Indeed, the Ricci scalar is (see \ref{s:calculations} for the computation)
\begin{equation} \label{Kretschmann&Ricci}
R=-\frac{24 m^2 b^2 r^2 \left(r^{3} - 4 m b^{2}\right)}{{\left(a^{2} \cos\left({\theta}\right)^{2} + r^{2}\right)} {\left(r^{3} + 2 m b^{2}\right)}^{3}} ,
\end{equation}
which is singular in the entire hypersurface defined by $r=-(2mb^2)^{1/3}$.
Similarly, the Kretschmann scalar
$K := R_{\mu\nu\rho\sigma} R^{\mu\nu\rho\sigma}$ diverges at the same
value of $r$ (cf. Fig.~\ref{R&K_plots_H}).
We conclude that the rotating generalization (\ref{metricKerrmodified})-(\ref{e:M_r_BambiModesto}) of Hayward's metric does not describe a regular
black hole.


\subsection{The rotating  Hayward metric extended to $r<0$}
\subsubsection{Metric}
\leavevmode \par

Following a prescription applied by Torres \cite{Torres:2017}
to rotating regular black boles arising from quantum gravity
consideration, we define the metric tensor in all
$\mathscr{M}= \mathbb{R}^2\times\mathbb{S}^2$ by
\begin{equation} \label{metric_improved_Hayward}
\begin{aligned}
\dd s^2=&-\left(1-\frac{2rM(r)}{\Sigma} \right)\dd t^2-\frac{4arM(r)\sin^2 \theta}{\Sigma}\, \dd t \, \dd\phi + \frac{\Sigma}{\Delta} \, \dd r^2+\Sigma \, \dd\theta^2 \\
&+\sin^2 \theta \left(r^2+a^2+\frac{2a^2rM(r)\sin^2\theta}{\Sigma} \right) \dd \phi^2,
\end{aligned}
\end{equation}
with
\begin{equation}
\label{e:M_r_Torres}
\begin{aligned}
 & \Sigma :=r^2+a^2\cos^2 \theta, \quad
 \Delta :=r^2-2M(r)r+a^2, \\
 & M(r) :=\displaystyle m \frac{|r|^3}{|r|^3+2mb^2}.
\end{aligned}
\end{equation}
The difference with Bambi-Modesto's metric (\ref{metricKerrmodified})-(\ref{e:M_r_BambiModesto}) lies only in the replacement of $r$ by $|r|$ in
the function $M(r)$. This is motivated by the expression of $M(r)$ in Torres' work
\cite{Torres:2017}:
\begin{equation}
    M(r)_{\rm Torres} = m \frac{|r|^3}{|r|^3 + \tilde\omega(|r| + \gamma m)},
\end{equation}
where $\tilde\omega$ and $\gamma$ are two constants.

\subsubsection{The hypersurface $r=0$}
\leavevmode \par

In $\mathscr{M}= \mathbb{R}^2\times\mathbb{S}^2$, the
hypersurface $r=0$ is
a 3-dimensional cylinder $\mathscr{T}_0 = \mathbb{R}\times\mathbb{S}^2$, spanned by
the coordinates $(t,\theta,\phi)$, which we call the \emph{throat},
as in the Kerr case \cite{ONeill:1995}. The metric
induced on $\mathscr{T}_0$  by the spacetime metric
(\ref{metric_improved_Hayward})-(\ref{e:M_r_Torres}) is
\begin{equation}
\label{e:metric_throat}
\dd\sigma^2=-\dd t^2+a^2\cos^2 \theta \, \dd\theta^2 +a^2 \sin^2 \theta \,
    \dd \phi^2.
\end{equation}
We may then split $\mathscr{T}_0$ into three components:
$\mathscr{T}_0 = \mathscr{T}_0^+ \cup \mathscr{R} \cup \mathscr{T}_0^-$,
where $\mathscr{T}_0^+$ is the Northern hemisphere $0\leq\theta<\pi/2$
times (Cartesian product) $\mathbb{R}$,
$\mathscr{R}$ is the equatorial ring $\theta=\pi/2$ times $\mathbb{R}$
and $\mathscr{T}_0^-$ is the Southern hemisphere
$\pi/2<\theta\leq\pi$ times $\mathbb{R}$.
Introducing in $\mathscr{T}_0^+$ or $\mathscr{T}_0^-$ the coordinates
\begin{equation}
\left\{
\begin{array}{ll}
X&=a \sin \theta \cos \phi, \\
Y&=a \sin \theta \sin \phi
\end{array}
\right.
\qquad X^2 + Y^2 \leq a^2 ,
\end{equation}
the line element (\ref{e:metric_throat}) reduces to
\begin{equation}
\dd\sigma^2=-\dd t^2 + \dd X^2+\dd Y^2.
\end{equation}
We recognize a 3-dimensional Minkowskian metric and conclude that,
as long as $a\not = 0$, the throat $\mathscr{T}_0$ comprises two flat open disks
of radius $a$ times $\mathbb{R}$: $\mathscr{T}_0^+$ and $\mathscr{T}_0^-$.
Moreover, from the signature of (\ref{e:metric_throat}), it appears that the
throat is timelike; it is therefore a 2-way membrane, i.e. it can be crossed by particles from the region $r>0$ to the region $r<0$, in the reverse way as well.

\subsubsection{Regularity}
\leavevmode \par

\begin{figure}[ht]
\hspace{-3cm}
  \begin{subfigure}[b]{0.75\linewidth}
    \centering
    \includegraphics[width=0.8\linewidth]{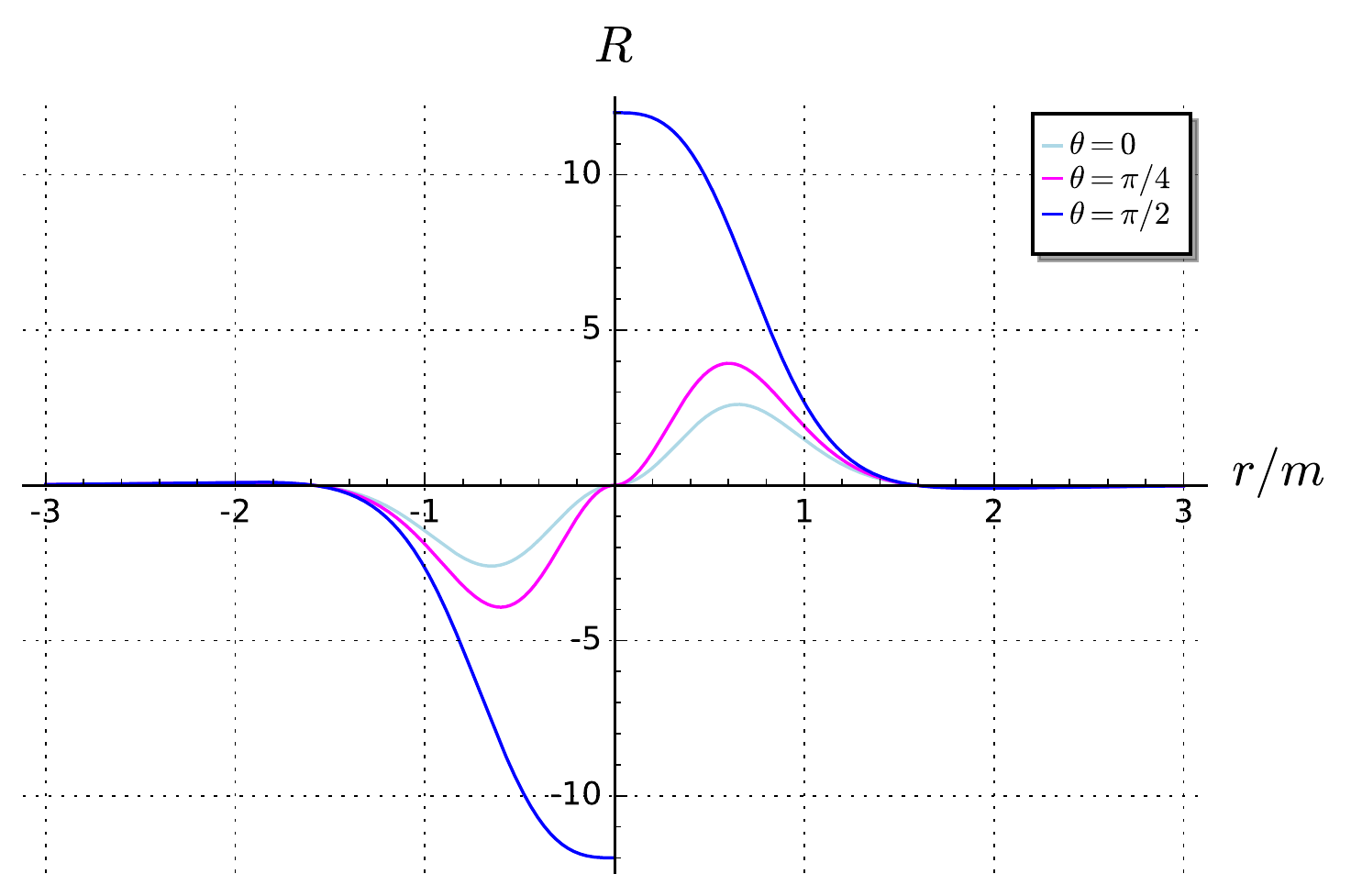}
    \caption{Ricci scalar}
    \label{R_plot_HT}
    \vspace{4ex}
  \end{subfigure}
\hspace{-2.5cm}
  \begin{subfigure}[b]{0.75\linewidth}
    \centering
    \includegraphics[width=0.8\linewidth]{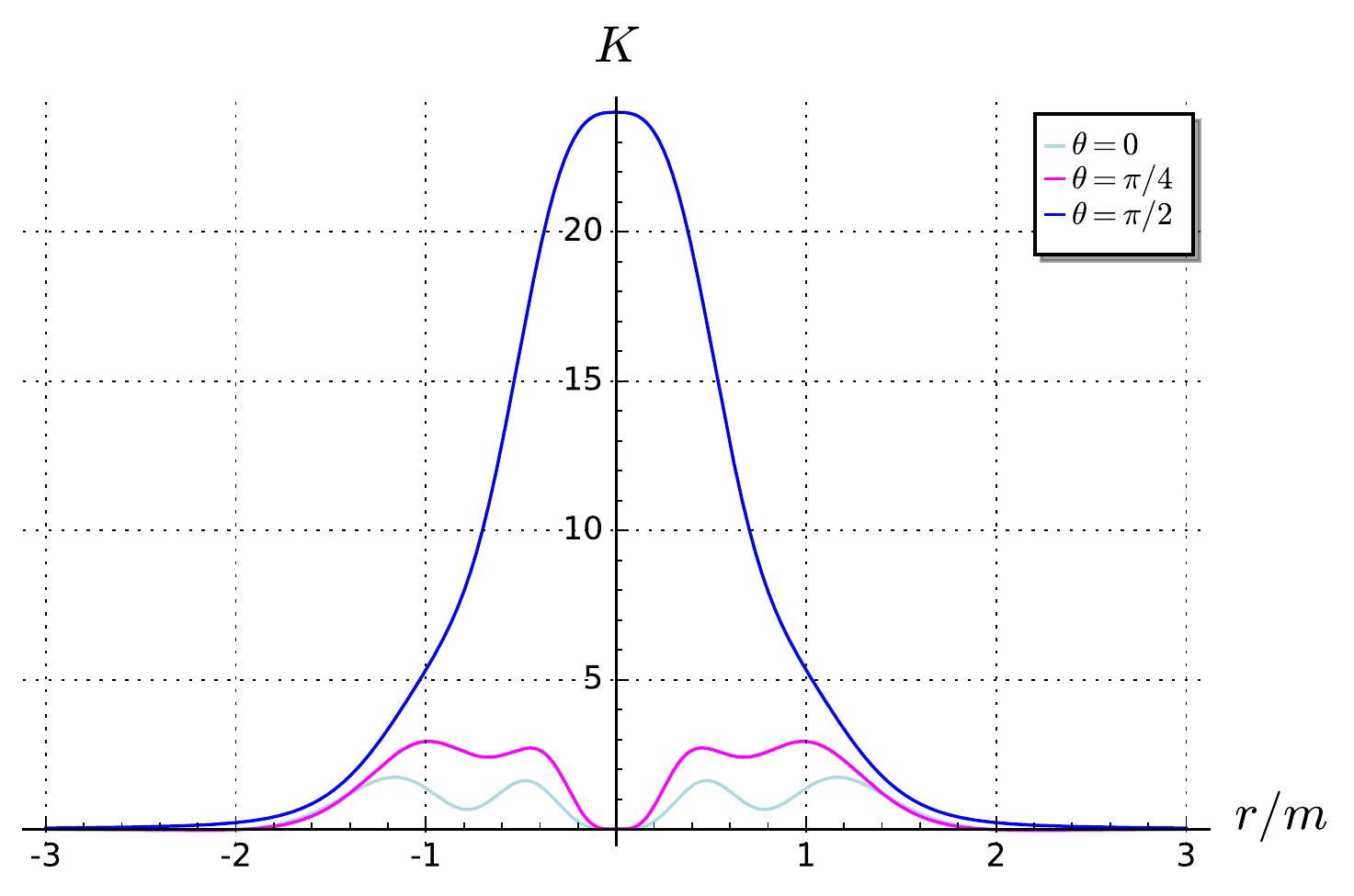}
    \caption{Kretschmann scalar}
    \label{K_plot_HT}
    \vspace{4ex}
  \end{subfigure}
  \vspace{-1.5cm}
  \caption{\footnotesize
Ricci scalar (a) (in units of $m^{-2}$) and Kretschmann scalar (b) (in units of $m^{-4}$) of the improved rotating Hayward metric (\ref{metric_improved_Hayward})-(\ref{e:M_r_Torres}) with $a/m=0.9$ and $b/m=1$
as a function of $r$ for $\theta=0$, $\pi/4$ and $\pi/2$.}
\label{}
\end{figure}

\begin{figure}[ht]
\hspace{-3cm}
  \begin{subfigure}[b]{0.75\linewidth}
    \centering
    \includegraphics[width=0.8\linewidth]{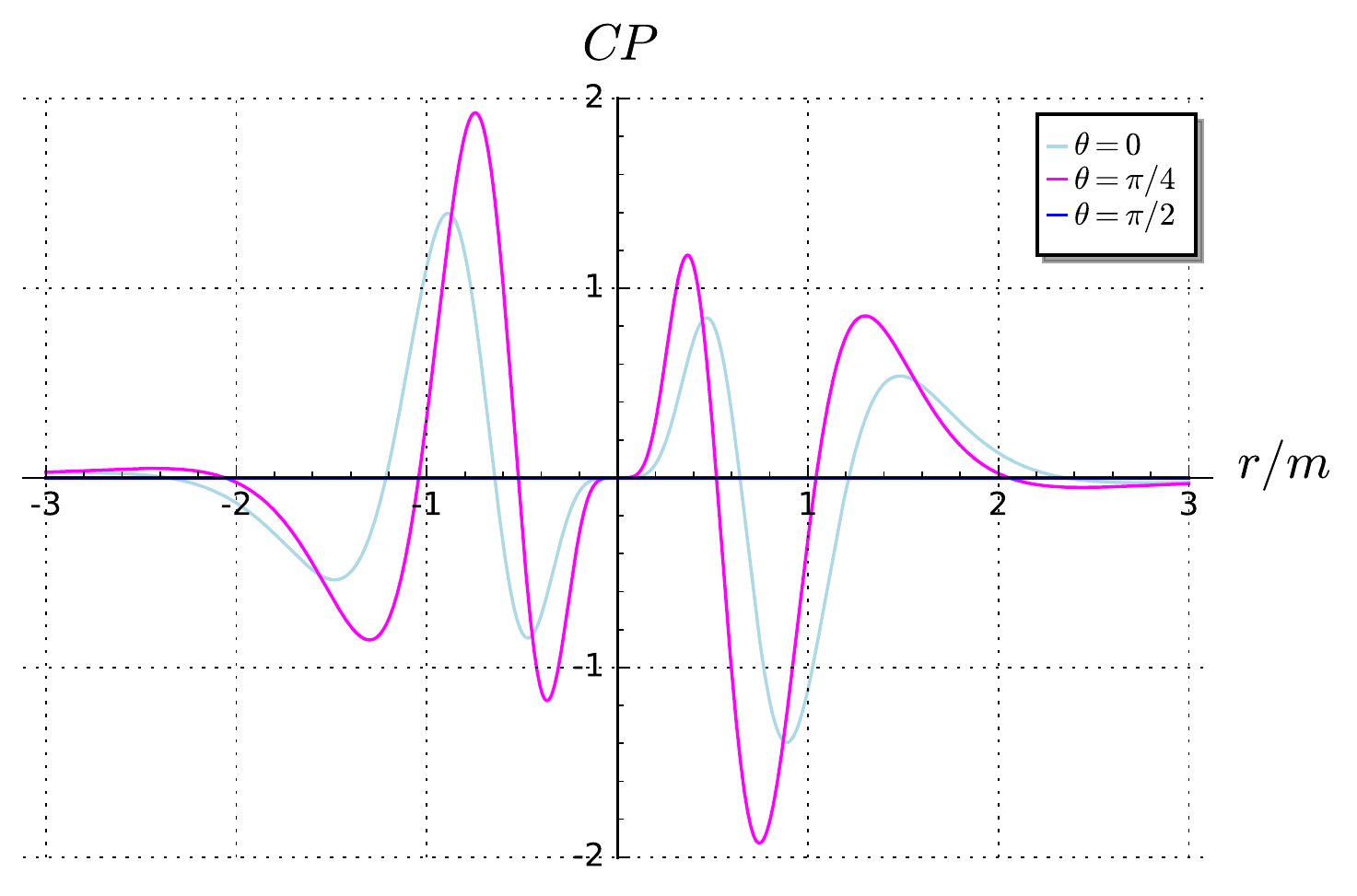}
    \caption{Chern-Pontryagin scalar}
    \label{CP_plot_HT}
    \vspace{4ex}
  \end{subfigure}
\hspace{-2.5cm}
  \begin{subfigure}[b]{0.75\linewidth}
    \centering
    \includegraphics[width=0.8\linewidth]{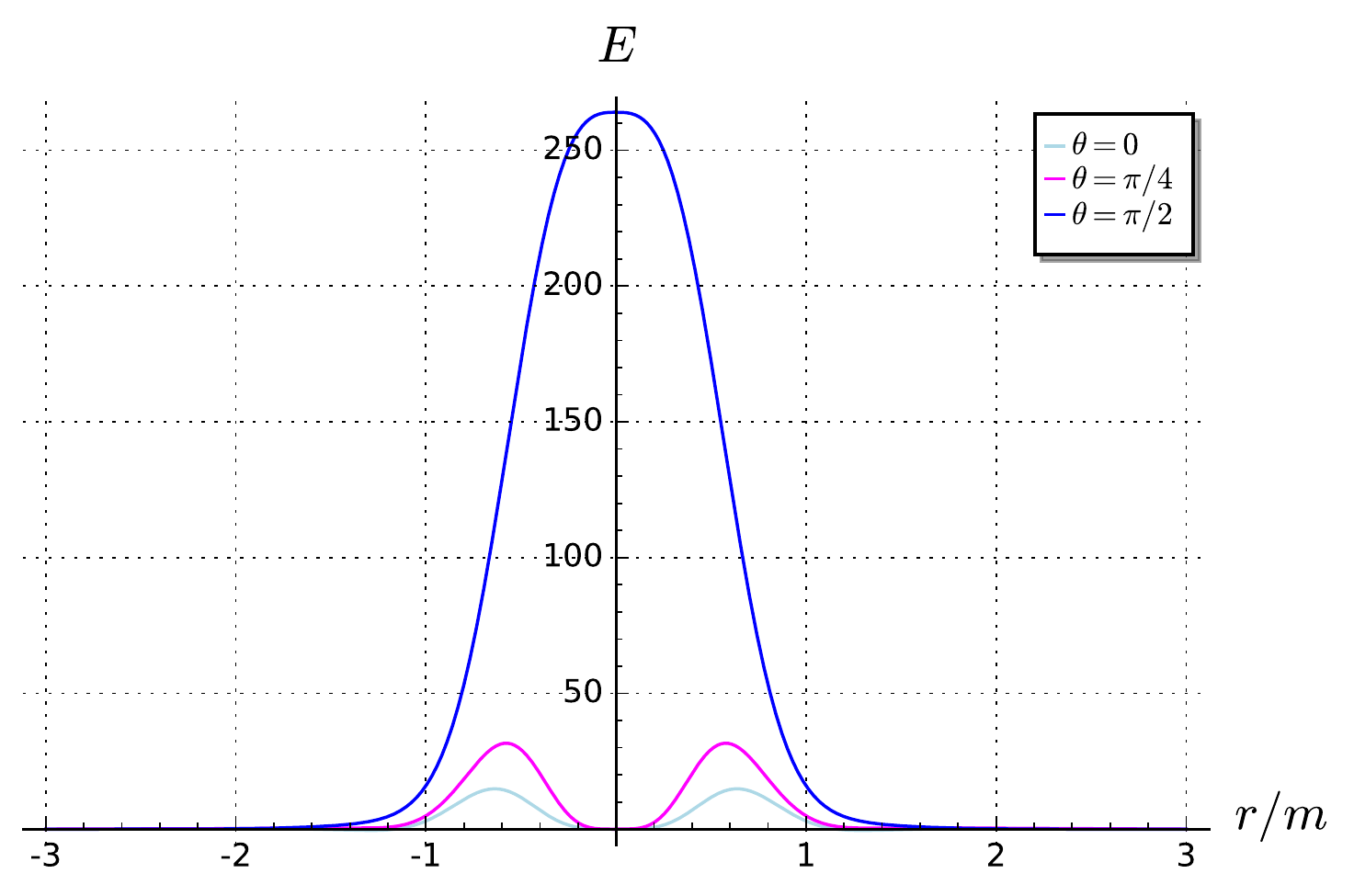}
    \caption{Euler scalar}
    \label{E_plot_HT}
    \vspace{4ex}
  \end{subfigure}
  \vspace{-1.5cm}
\caption{\footnotesize
Chern-Pontryagin scalar (a) and Euler scalar (b) (in units of $m^{-4}$) of the improved rotating Hayward metric (\ref{metric_improved_Hayward})-(\ref{e:M_r_Torres}) with $a/m=0.9$ and $b/m=1$
as a function of $r$ for $\theta=0$, $\pi/4$ and $\pi/2$.}
\label{}
\end{figure}

The metric (\ref{metric_improved_Hayward})-(\ref{e:M_r_Torres})
has no curvature singularity. This can be seen on Figs.~\ref{R_plot_HT},
\ref{K_plot_HT}, \ref{CP_plot_HT} and \ref{E_plot_HT}, where the Ricci scalar $R$, Kretschmann scalar $K$, Chern-Pontryagin scalar $CP$ and Euler scalar $E$ are plotted for $a/m=0.9$, $b/m=1$ and various values of $\theta$ (see \ref{s:calculations} for details). The curvature scalars
remain finite, although the Ricci scalar is discontinuous at the equatorial
ring $r=0$ and $\theta=\pi/2$.

\subsubsection{Horizons}
\leavevmode \par

\begin{figure}[!h]
\begin{center}
\includegraphics[scale=0.8]{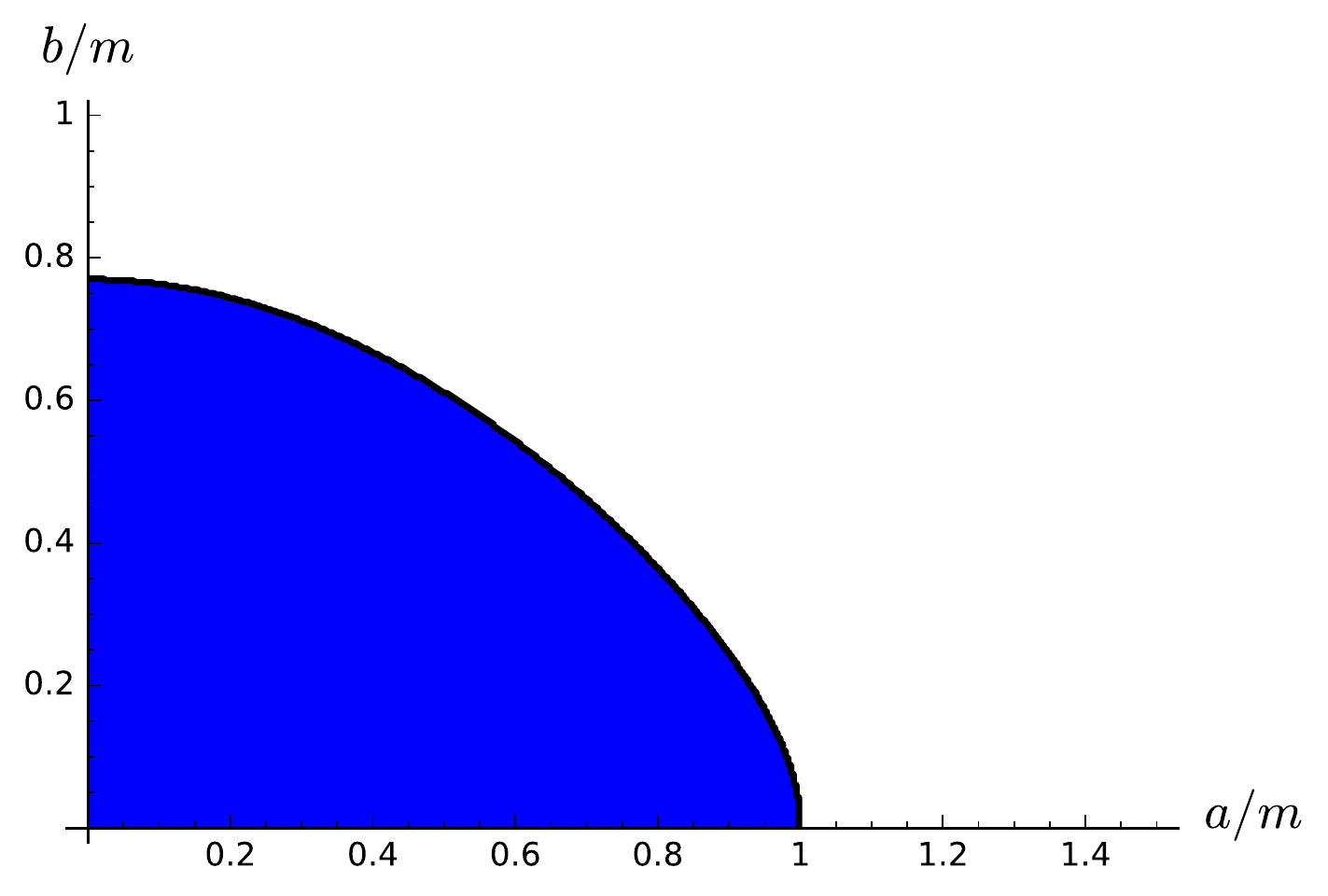}
\caption[]{\label{Horizon_a_b} \footnotesize
 Region of existence of one (black line) or two (in blue) Killing horizon(s), depending on the parameters $a$ and $b$.}
 \end{center}
\end{figure}

In the context of the stationary metric (\ref{metric_improved_Hayward}), the trapping horizons, which identify to Killing horizons, are the null hypersurfaces where the expansion of a congruence of null outgoing geodesics vanishes. This condition reduces to
\begin{equation} \label{presence_horizons}
\Delta = r^2-2M(r)r+a^2=0.
\end{equation}
This equation admits some real solutions depending on the values of the parameters $a$ and $b$. We will call \emph{event horizon} the outermost Killing horizon, which corresponds to the biggest value of the radial coordinate among the solutions of (\ref{presence_horizons}). The region of existence of Killing horizons is depicted on Fig. \ref{Horizon_a_b}.

When $b=0$ one recovers the Kerr case: two horizons exist for values of $a$ ranging from $a=0$ to $a=m$, the latter value corresponding to the extremal Kerr black hole, where the two horizons coincide. The most interesting cases with horizons are the metrics well different from Kerr ($b=0$) and Hayward ($a=0$) ones, for instance the metric with $a=b=0.5\,m$. The image of such configurations, computed using the ray-tracing code\footnote{Freely available at
\url{http://gyoto.obspm.fr}} \gy~\cite{Vincent_et_al:2011}, will be discussed in Sec. \ref{improved_Hawyard_horizons}. \\

In the absence of horizon (hence of trapped region), the spacetime can no longer be qualified of a regular rotating black hole. That is why we call it a \emph{naked rotating wormhole}. Indeed, the wormhole whose throat is located at $r=0$, which is also present in Kerr's case (with a singularity), is no longer hidden by any horizon. Photons can even go through the throat and come back to the observer, as will be shown in Sec.~\ref{Naked rotating wormhole}.

\subsubsection{Causality}
\leavevmode \par

The Kerr spacetime possesses a well-known acausal region, the Carter time machine \cite{Carter:1968}. In this region, the Killing vector $\eta=\partial_\phi$ is timelike, giving birth to closed timelike curves. However the whole spacetime does not become acausal thanks to the presence of an event horizon: the particles
that are able to move backward in time are trapped inside the black hole. \\

\begin{figure}[!h]
\begin{center}
\includegraphics[scale=0.8]{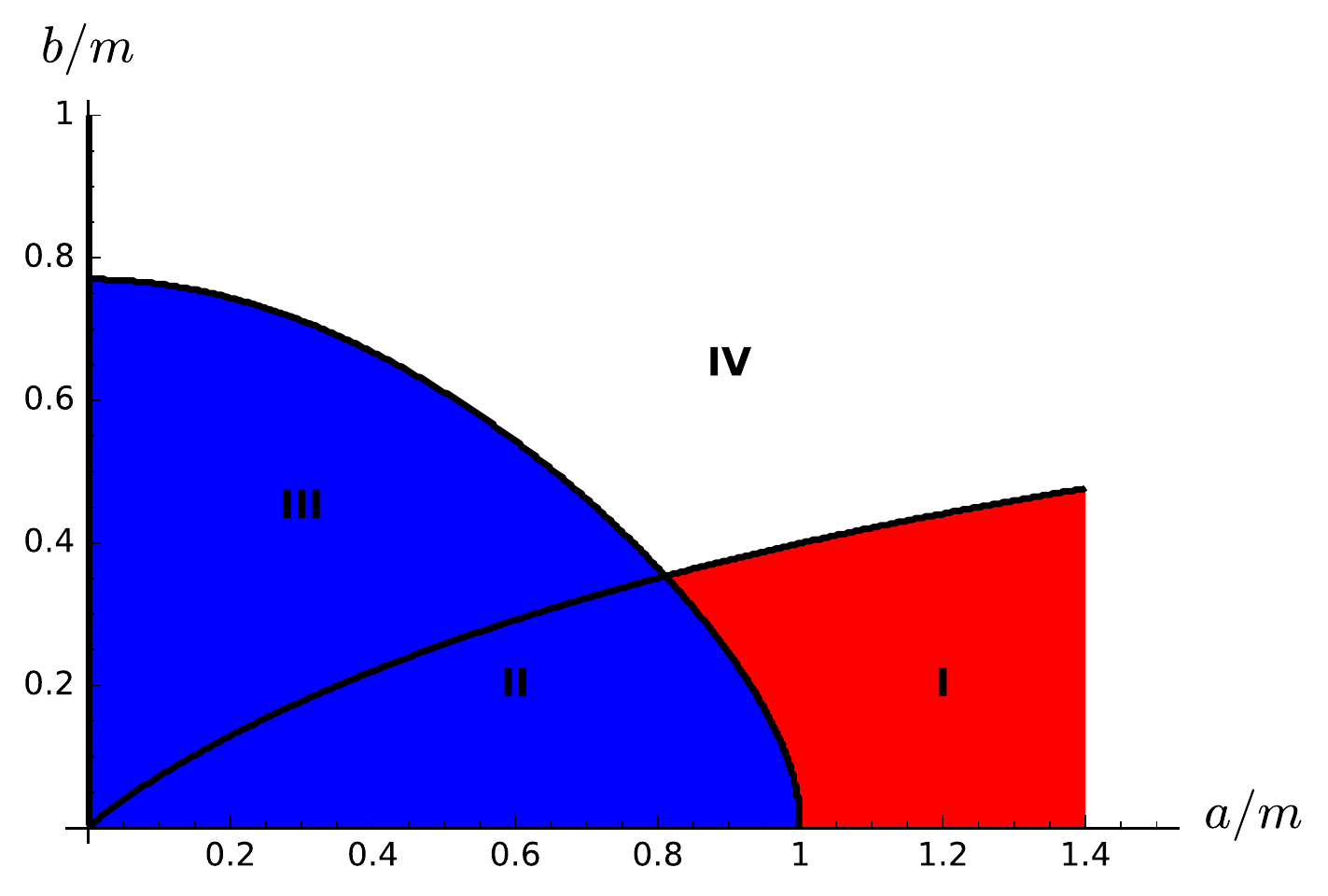}
\caption[]{\label{Causality_Horizon_a_b} \footnotesize
 Regions of existence of an event horizon (in blue) and of negative $g_{\phi \phi}$ for $\theta=\pi/2$ in the absence of horizons (in red), depending on the parameters $a$ and $b$.}
 \end{center}
\end{figure}

Considering now the rotating Hayward metric extended to $r<0$, one has to check whether $\eta$ can become timelike even in the absence of horizons, in which case the whole spacetime would be acausal. In view of (\ref{metric_improved_Hayward}), one has
\begin{equation}
\eta \cdot \eta=g_{\phi \phi}=\left(r^2+a^2+\frac{2a^2M(r)r\sin^2\theta}{r^2+a^2\cos^2\theta} \right) \sin^2 \theta,
\end{equation}
so that
\begin{equation} \label{eta_timelike}
\eta \hspace{2mm} \mbox{timelike} \hspace{2mm} \Leftrightarrow (r^2+a^2)(r^2+a^2\cos^2\theta) + 2a^2M(r)r\sin^2\theta<0 .
\end{equation}
The only negative contribution in the left-hand side of (\ref{eta_timelike}) comes from the second term, when $r<0$. It reaches a minimum for $\theta=\pi/2$. Fig. \ref{Causality_Horizon_a_b} shows that there exists a red region (region I) for which $g_{\phi \phi}<0$ while no event horizon is present. The parameters $a$ and $b$ associated with such a region thus correspond to acausal spacetimes, which we will not deal with in this paper.

Region II also has causality issues, but these are hidden behind an event horizon. Regions III and IV are totally free of closed timelike curves, the latter is also devoid of any event horizon and represents a naked rotating wormhole.

\subsubsection{Energy conditions}
\leavevmode \par

The existence of horizons, hence of trapped surfaces, along with the absence of singularity, questions the hypotheses of Penrose's singularity theorem.
As mentionned in Sec.~\ref{s:intro}, both Bardeen and Hayward nonrotating metrics
fulfill the weak energy condition and circumvent the original Penrose theorem
\cite{Penrose:1965} by the lack of a Cauchy surface. In an improved version
of the singularity theorem, by Hawking and Penrose (1970) \cite{HawkingP:1970,HawkingE:1973},
the hypothesis of existence of a Cauchy surface is relaxed, at the price
of replacing the weak energy condition by the strong one. This version is
still compatible with Bardeen's and Hayward's regular black holes because
both violate the strong energy condition.

In the rotating case, it has been shown by Torres \cite{Torres:2017}
that any metric of the type (\ref{metric_improved_Hayward}) with $a\not=0$
violates the
weak energy condition in all the region $r<0$ as soon as $M'(r)<0$ there.
This is the case for our choice (\ref{e:M_r_Torres}) for $M(r)$.

\begin{figure}[!h]
\begin{center}
\includegraphics[scale=0.8]{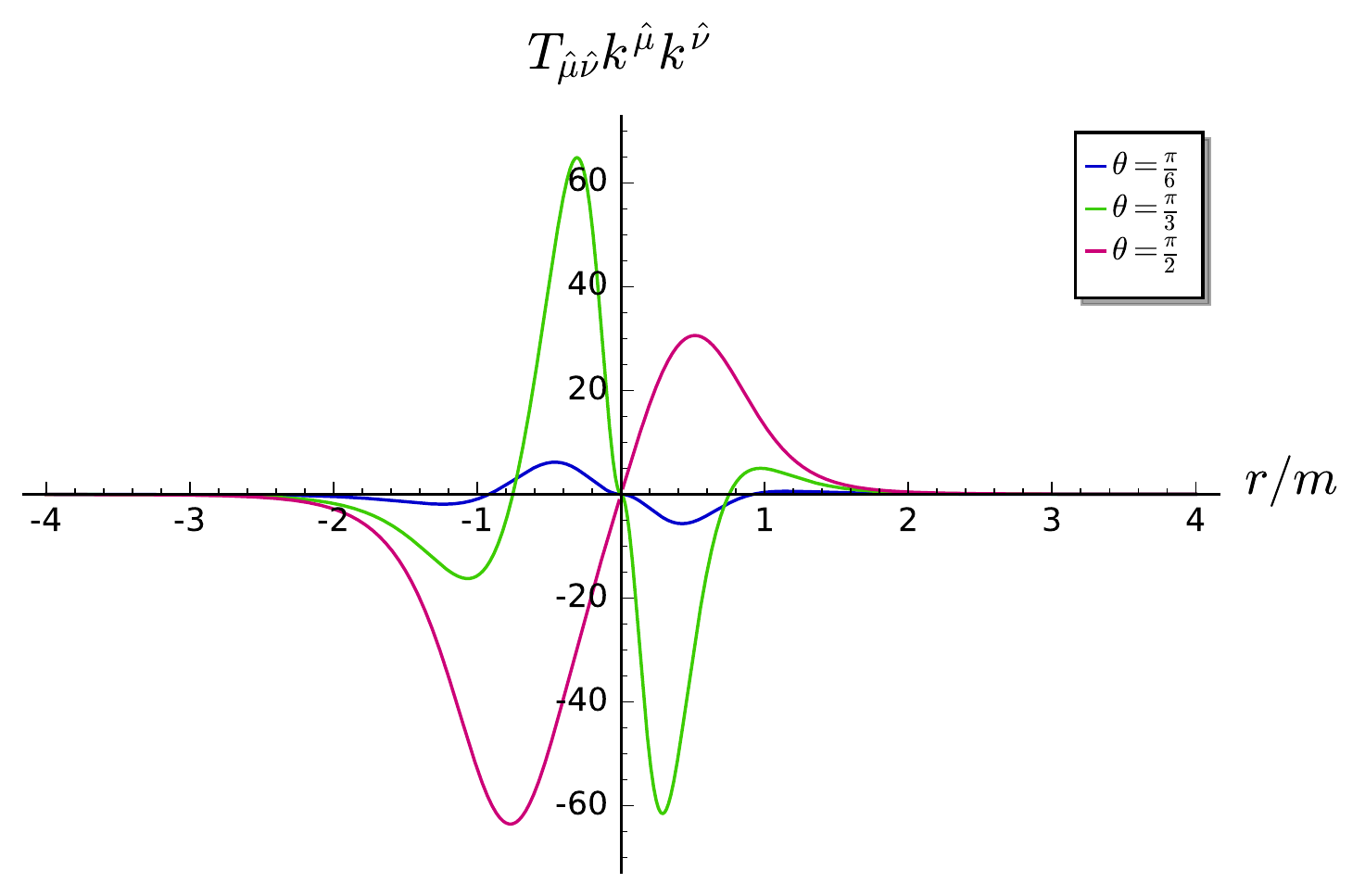}
\caption[]{\label{NEC_a09_b1} \footnotesize
$T_{\hat{\mu}\hat{\nu}}k^{\hat{\mu}}k^{\hat{\nu}}$ as a function of $r$ for $\theta=\pi/6,\pi/3,\pi/2$ and $a=0.9, b=1$. The NEC is violated when any of the curves goes below zero.}
 \end{center}
\end{figure}

Here, we investigate the violation of the weakest of all energy conditions, the \emph{null energy condition (NEC)}. It is the weakest condition in the sense that its violation also implies the violation of the weak, strong and dominant energy conditions. For any null vector $k^{\mu}$ the NEC reads
\begin{equation}
T_{\mu \nu}k^{\mu}k^{\nu} \geq 0.
\end{equation}
In order to compute this scalar we switch to the locally nonrotating frame which diagonalizes the metric \cite{Bardeen:1972}. Its  basis is such that $e_{\hat{\mu}} \cdot e_{\hat{\nu}}=\eta_{\hat{\mu}\hat{\nu}}$.
The dual cobasis at each point $(t,r,\theta,\phi)$ reads
\begin{equation}
\begin{aligned}
e^{(t)}&=\sqrt{\frac{\Sigma \Delta}{A}} \,\dd t, \\
e^{(r)}&=\sqrt{\frac{\Sigma}{\Delta}} \,\dd r, \\
e^{(\theta)}&=\sqrt{\Sigma}\,\dd\theta, \\
e^{(\phi)}&=-\frac{2M(r)ar\sin \theta}{\sqrt{\Sigma A}}\,\dd t+\sqrt{\frac{A}{\Sigma}} \sin \theta \,\dd\phi,
\end{aligned}
\end{equation}
with
\begin{equation}
A:=(r^2+a^2)^2-a^2\Delta \sin^2 \theta.
\end{equation}
Solving Einstein's equations ``in reverse'', we obtain $T_{\hat{\mu}\hat{\nu}}k^{\hat{\mu}}k^{\hat{\nu}}=G_{\hat{\mu}\hat{\nu}}k^{\hat{\mu}}k^{\hat{\nu}}/8\pi$. This  effective energy density is plotted in Fig.~\ref{NEC_a09_b1} in the case $a=0.9$, $b=1$ (see \ref{s:calculations} for details). One can see that the NEC is violated from near the centre up to $r\rightarrow -\infty$.

\section{Numerical study of the regular rotating Hayward metric with \gy}
\label{section 3}
\subsection{Regular rotating Hayward black hole} \label{improved_Hawyard_horizons}

This section aims at discussing the differences in ray-traced images between the regular rotating Hayward black hole and the standard Kerr black hole. \\

We now use the ray-tracing code \gy to obtain images from an accretion torus surrounding the two black holes (see \ref{Gyoto_plugin} for details). Another approach could consist in computing analytically the contour of the shadow based on the formula developed by Tsukamoto \cite{tsukamoto_2018}, but here we opt for a numerical computation in an astrophysical context instead. The set-up is composed of an accretion torus  identical to the one presented in Ref.~\cite{Vincent_et_al:2015}. It is a magnetized
optically thin torus, with angular momentum $l=4m$, inner radius $r_\mathrm{inner}=8.3m$, where we take $m$ to be the mass of Sgr A* ($m=4.31 \times 10^6 \hspace{2mm} M_{\odot}$). The outer radius varies according to the value of the spin parameter, it is for instance  $r_\mathrm{outer}=30m$ for $a=0.9$.
The central temperature is of $T=5.3\times10^{10}$~K and the central electron number density $n_e = 6.3\times 10^6\,\mathrm{cm}^{-3}$.
We compute the thermal synchrotron radiation emitted in the millimeter band by this torus, the ray-traced photons are observed at a frequency of 230 GHz. Such an accretion flow
was shown (see~\cite{Vincent_et_al:2015}) to reproduce well the millimeter spectral data of Sgr~A*, as well as the constraints on the size
of the emitting region imposed by early EHT data~\cite{Doeleman:2008}. The observer is located at a radial coordinate which corresponds to the distance between Earth and Sgr A*, and at an inclination (angle between the black hole rotation axis and the line of sight) of $\theta=90^{\circ}$. \\

In the case $a/m=0.5$, $b/m=0.5$, the outer horizon is located at $r_0 \approx 1.65m$. It is thus at a smaller value of the radial coordinate from the center than in the Kerr black hole case ($b/m=0$), where $r_+=m+\sqrt{m^2-a^2}\approx 1.87m$. Hence, for a given ADM mass, the black hole radius is smaller when $b \neq 0$.

\begin{figure}[ht]
\hspace{-2.5cm}
  \begin{subfigure}[b]{0.75\linewidth}
    \centering
    \includegraphics[width=0.85\linewidth]{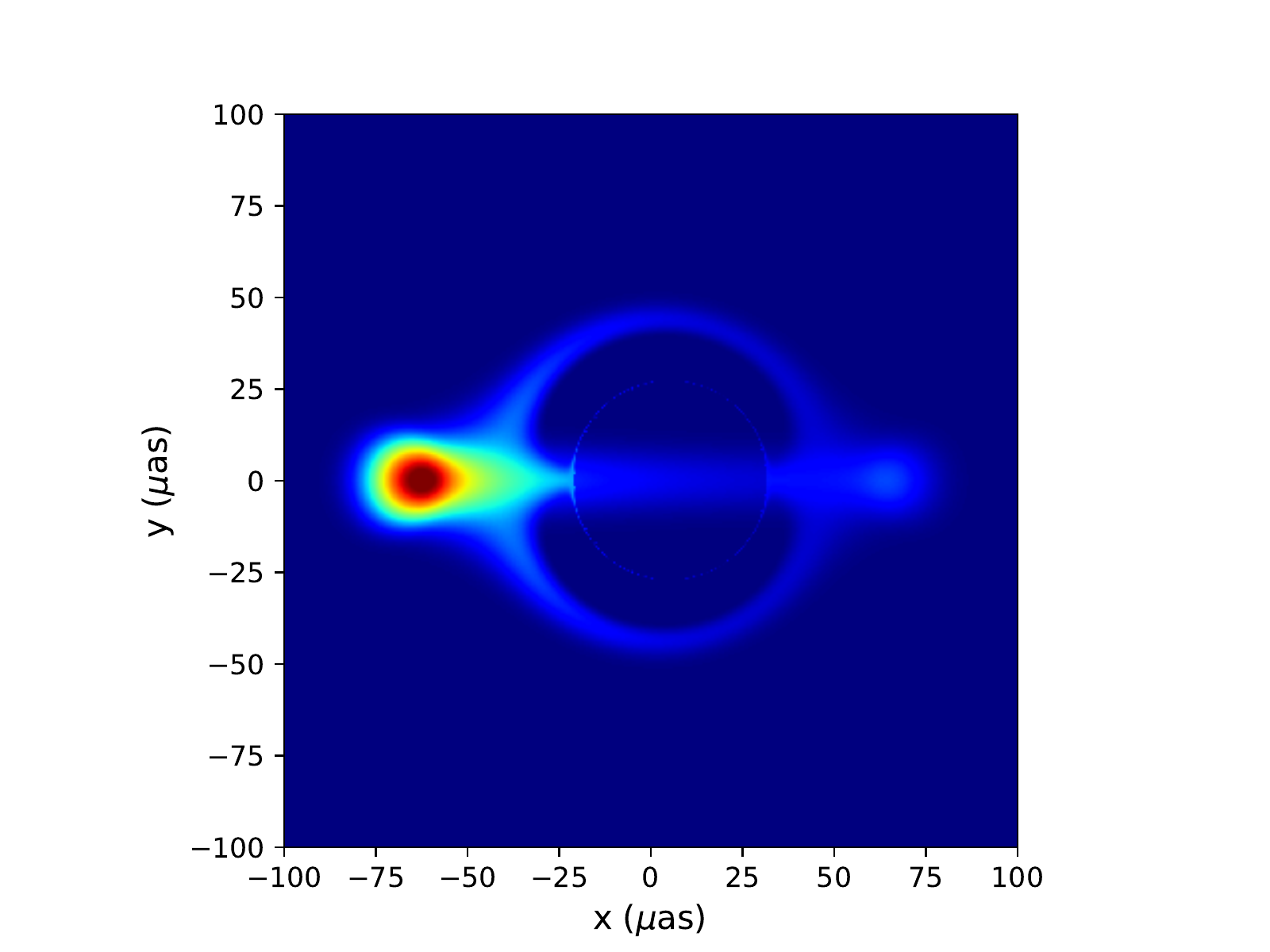}
    \caption{$a/m=0.5$, $b/m=0$}
    \label{fig6:a}
    \vspace{4ex}
  \end{subfigure}
\hspace{-3.8cm}
  \begin{subfigure}[b]{0.75\linewidth}
    \centering
    \includegraphics[width=0.85\linewidth]{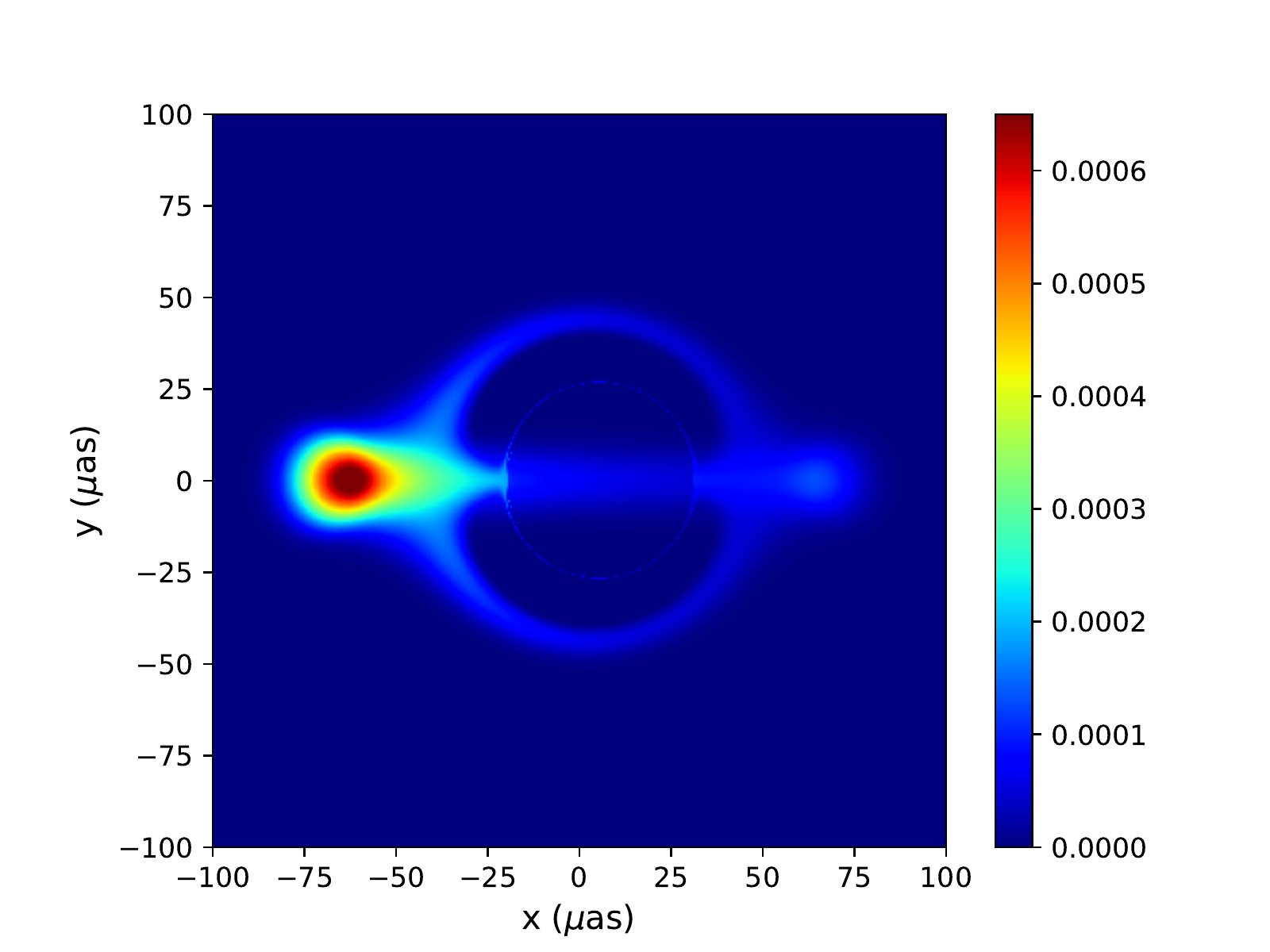}
    \caption{$a/m=0.5$, $b/m=0.5$}
    \label{fig6:b}
    \vspace{4ex}
  \end{subfigure}
  \vspace{-1.5cm}
  \caption{\footnotesize
Images of an accretion torus surrounding a Kerr black hole (a) and a regular rotating Hayward black hole (b), seen from a distance of $8.31$ kpc. The field of view is $200 \hspace{1mm} \mu\mbox{as}$ and the inclination $\theta=90^{\circ}$. The specific intensity $I_{\nu}$ is plotted in CGS units, as will be the case for the following images.}
\label{2BH with horizons}
\end{figure}

\begin{figure}[!h]
\begin{center}
\includegraphics[scale=0.8]{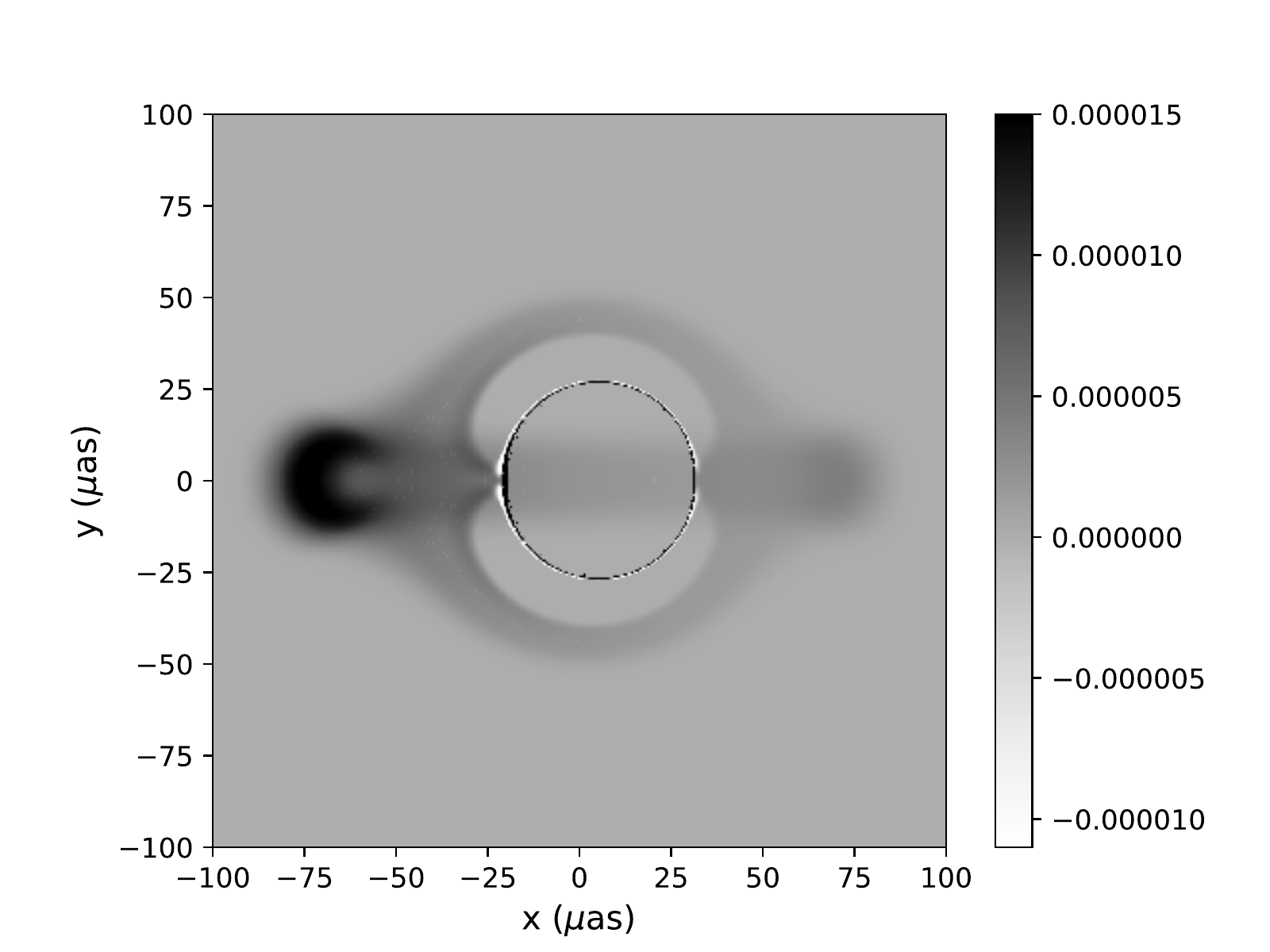}
\caption[]{\label{Difference_a05b05_a05b0}\footnotesize Difference of the images of Fig. \ref{2BH with horizons}. The lensing rings of the configurations $b/m=0.5$ (black) and $b/m=0$ (white) are visible at the centre.}
 \end{center}
\end{figure}

The millimeter images of these two black holes are visible on Fig. \ref{2BH with horizons}. On both panels, we observe the distorted primary image of the torus, that forms an Einstein ring. The very center of the image shows a thin lensing ring delineating the black hole shadow. The shadow is defined as the region in the observer’s sky comprising the directions of photons that asymptotically approach the event horizon in a backward ray-tracing computation \cite{VincentGourgoulhonHerdeiroRadu:2016}. The lensing ring is the secondary image of the torus. The higher-order images of the torus,
hardly visible on Fig.  \ref{2BH with horizons}, converge to the photon ring, which is the projection of the innermost photon orbit on the observer's sky. This photon orbit marks the innermost limit an approaching photon can visit without falling into the event horizon \cite{VincentGourgoulhonHerdeiroRadu:2016}. The differences between the millimeter images of the two black holes appear to be indistinguishable with the naked eye. However, substracting one image from another we can distinguish the two different lensing rings (Fig. \ref{Difference_a05b05_a05b0}). The difference of diameter between these two rings is about 2 $\mu$as ($\approx 3$\%). This difference is  out of reach for the observations in the foreseeable future, but we may  hope that a telescope would be able to measure the radius of the lensing ring in a far future, or equivalently the area of the shadow, and could thus discriminate between the two black holes for a given ADM mass.

It should be noted that we considered here a macroscopic value of $b$ ($b=0.5m$) when computing these images. The underlying assumption is that this parameter arises because of
some ``macroscopic'' energy-momentum tensor supporting this regular geometry,
similar to that arising from a nonlinear-electrodynamics magnetic monopole in the non-rotating case \cite{AyonG:2000,Fan:2017,Fan&Wang:2016}.
Had we assumed that the singularity was resolved by using $b$ as a Planckian cut-off, the difference between the images would have been invisible to any telescope even in the far future.

\subsection{Naked rotating wormhole}
\label{Naked rotating wormhole}

Let us now discuss the case of geometries without horizons described by the rotating Hayward metric extended to $r<0$ (\ref{metric_improved_Hayward}), that we call naked rotating wormholes. We will first describe the results obtained with \gy as well as their consequences, and then explain them by studying some relevant geodesics. \\

\begin{figure}[ht]
\hspace{-2.5cm}
  \begin{subfigure}[b]{0.75\linewidth}
    \centering
    \includegraphics[width=0.85\linewidth]{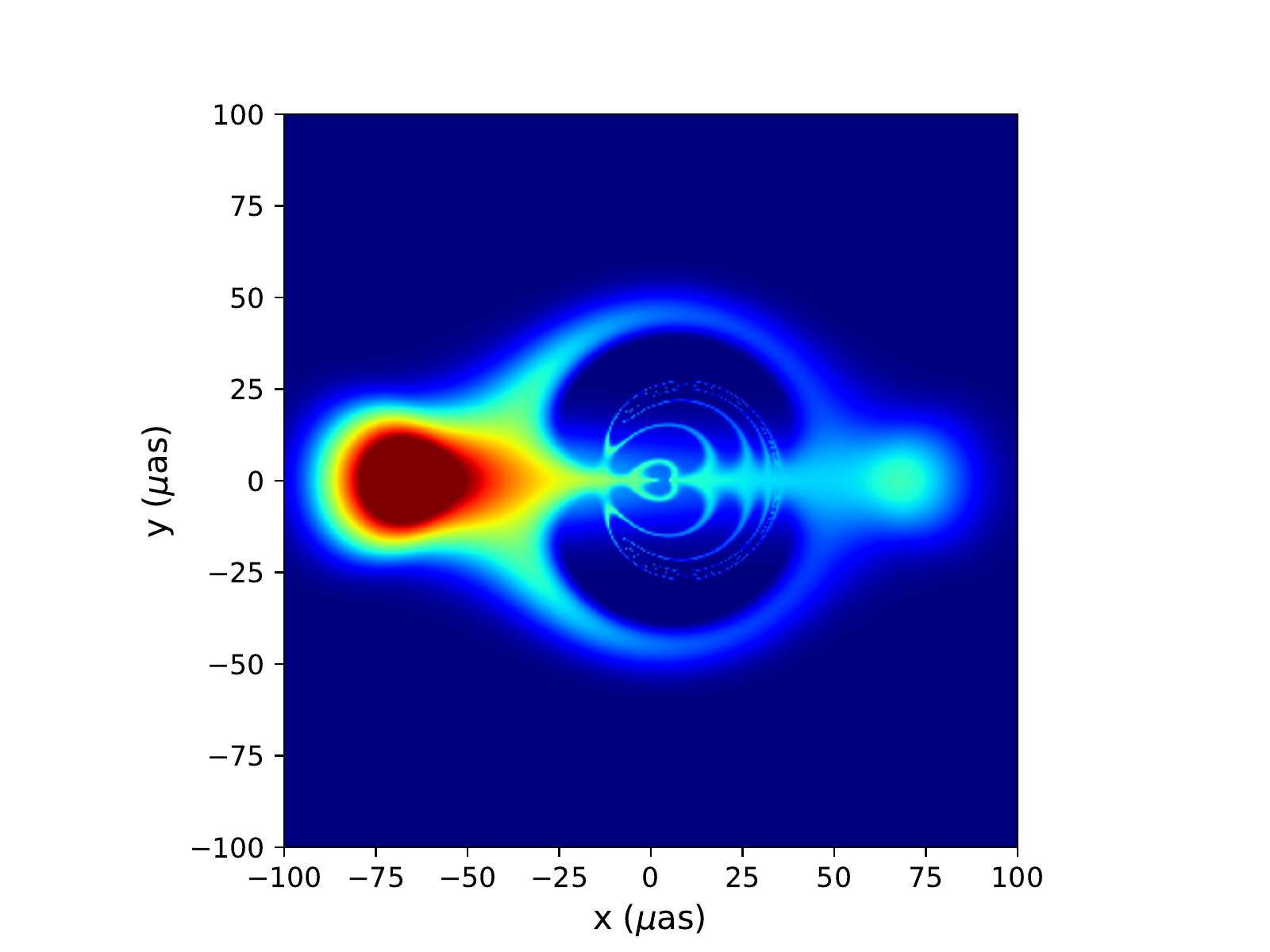}
    \caption{$a/m=0.9$, $b/m=0.4$}
    \label{fig8:a}
    \vspace{4ex}
  \end{subfigure}
\hspace{-4cm}
  \begin{subfigure}[b]{0.75\linewidth}
    \centering
    \includegraphics[width=0.85\linewidth]{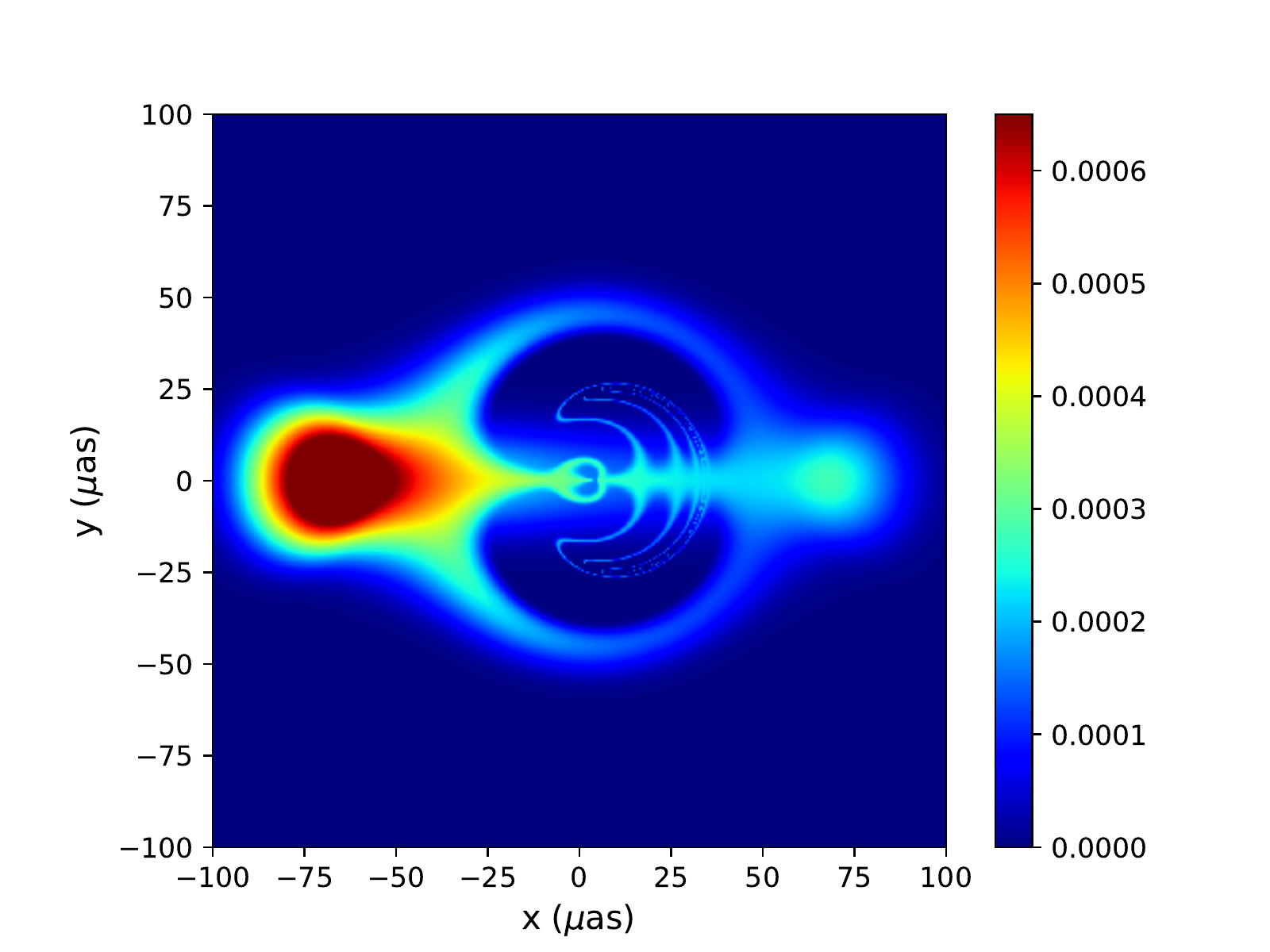}
    \caption{$a/m=0.9$, $b/m=0.7$}
    \label{fig8:b}
    \vspace{4ex}
  \end{subfigure}
  \vspace{-1.5cm}
  \caption{\footnotesize
 Images of an accretion torus surrounding a naked rotating wormhole with $a/m=0.9$, $b/m=0.4$ (a) and $b/m=0.7$ (b). The field of view is $200\; \mu{\rm as}$ and the inclination $\theta=90^{\circ}$.}
\label{no_shadow}
\end{figure}

The major difference in this configuration, with respect to the previous section, is the absence of horizons. Hence, the images obtained with this geometry do not contain any shadows, in the precise sense defined above. However, they do contain a central faint
region showing a mixture of low-flux regions and strongly lensed contours (of increasing order from left to right, see Fig. \ref{no_shadow}). The shape of these contours highly depends on the value of the parameter $b/m$. It it is thus very important to stress that observing such strongly lensed contours, which can look like a lensing ring  without a good enough resolution, does not imply the existence of an event horizon, just as in the case of boson stars \cite{Vincent_et_al:2016}. \\

The two panels of Fig.~\ref{no_shadow} are remarkably similar to the images of accretion tori
surrounding rotating boson stars (see the middle- and lower-right panels of Fig. 5 in Ref.~\cite{Vincent_et_al:2016}).
However, these spacetimes are completely different, the spacetimes analyzed here being naked rotating wormholes spacetimes, while boson stars are compact distributions of fundamental scalar fields. It is thus rather intriguing
that such very different spacetimes lead to images that are difficult to differentiate. Further studies would be necessary
in order to determine whether the distorted, hyper-lensed contours of Fig.~\ref{no_shadow} are general features
of spacetimes of compact object with no event horizon and no hard surface (i.e. different from neutron stars).

\begin{figure}[!h]
\centering
\begin{subsubcaption}
\begin{subfigure}{11cm}
\hspace{-2.1cm} \includegraphics[width=\linewidth]{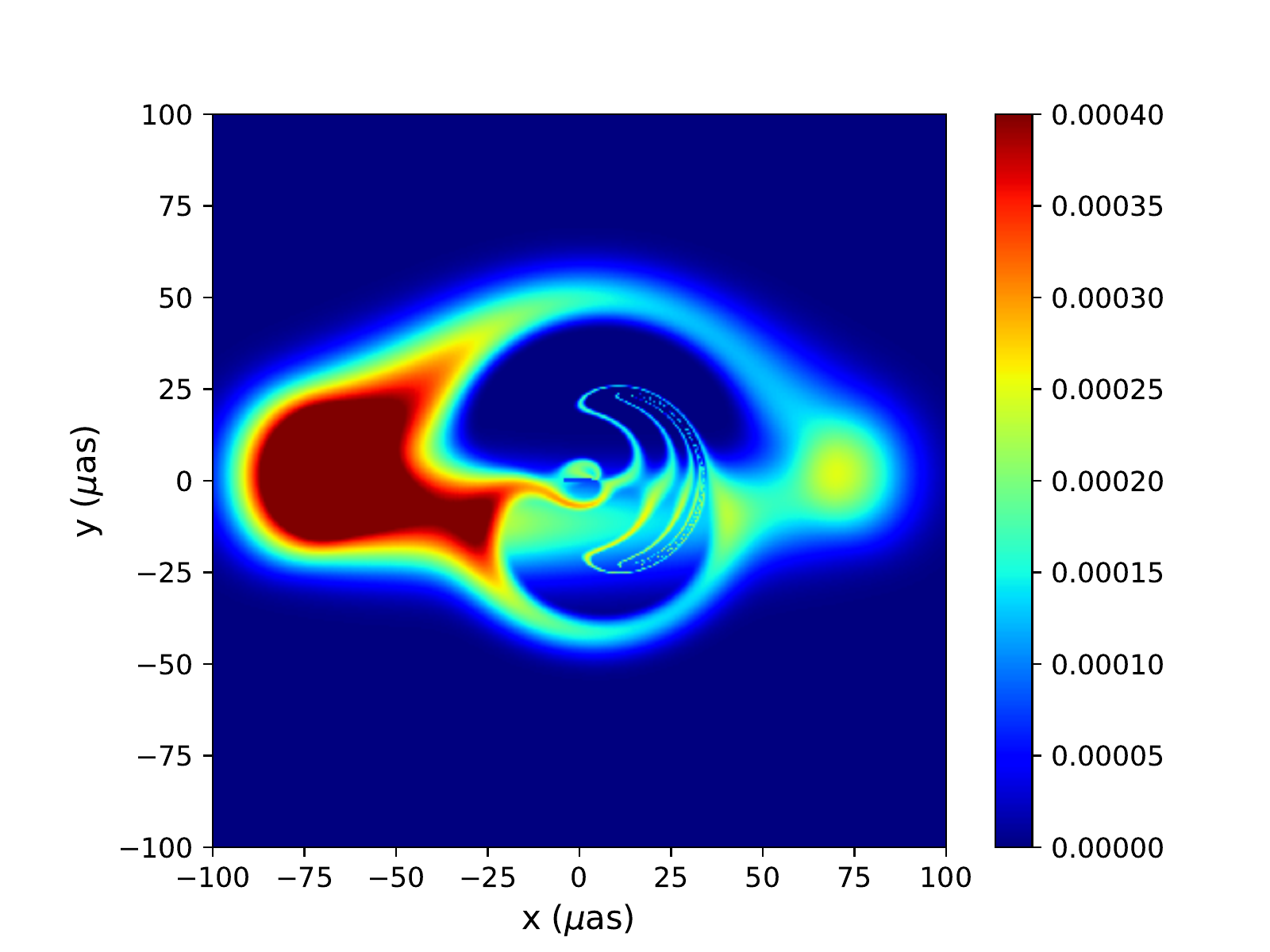}
\end{subfigure}\qquad
\hspace{-4cm}
\begin{subfigure}{11cm}
\includegraphics[width=\linewidth]{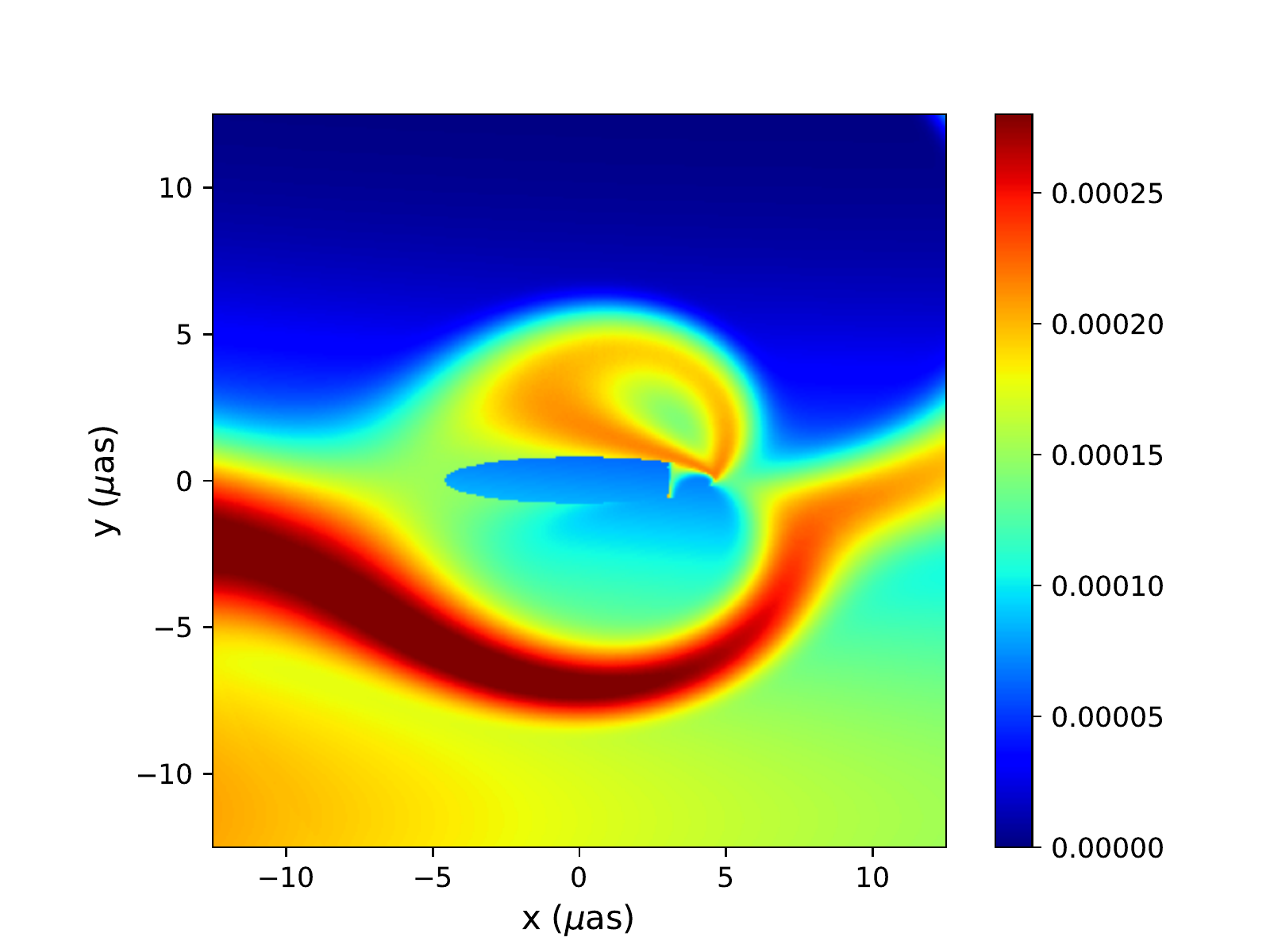}
\end{subfigure}
\end{subsubcaption}

\begin{subsubcaption}
\begin{subfigure}{11cm}
\hspace{-2.1cm} \includegraphics[width=\linewidth]{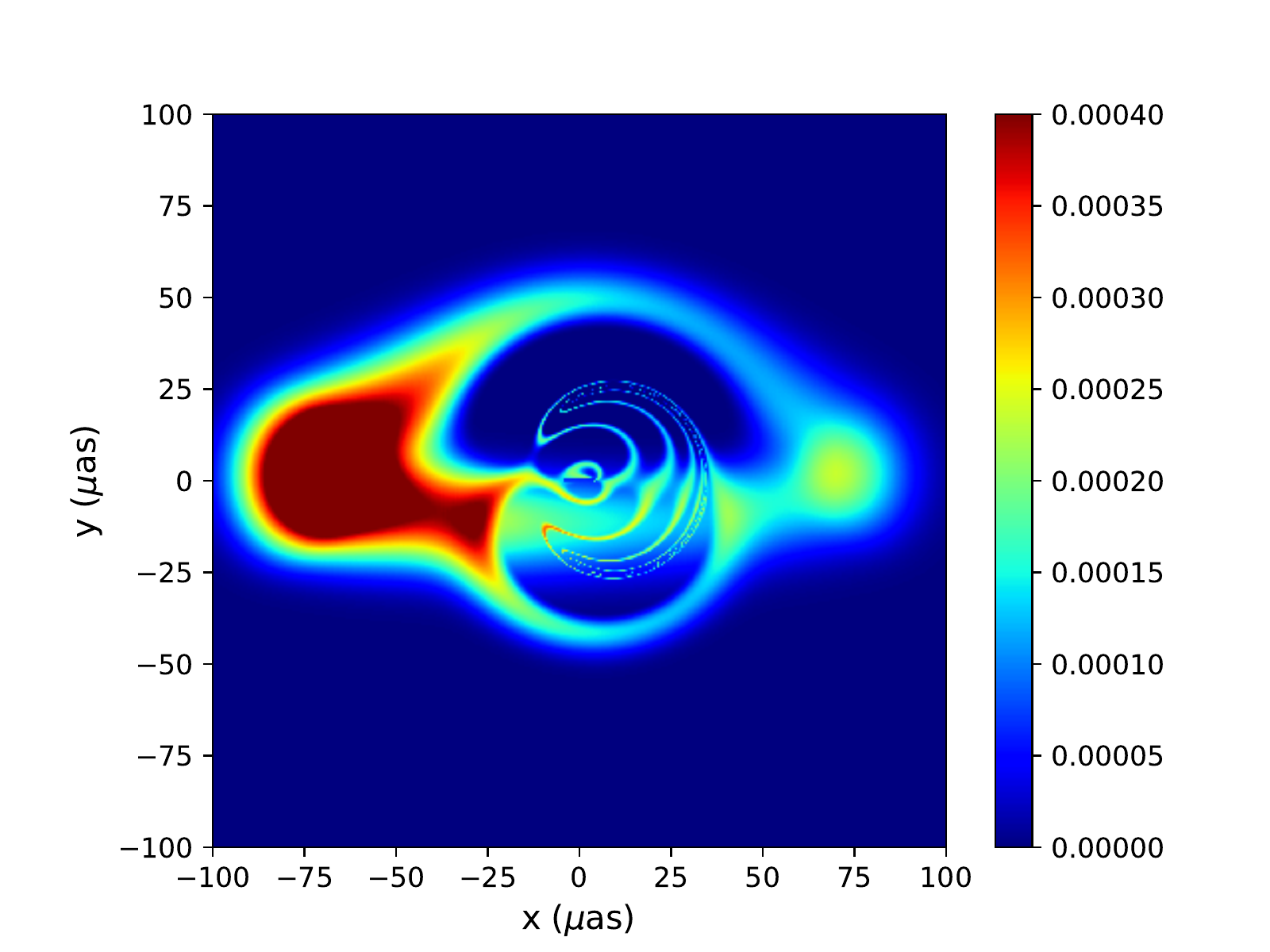}
\end{subfigure}\qquad
\hspace{-4cm}
\begin{subfigure}{11cm}
\includegraphics[width=\linewidth]{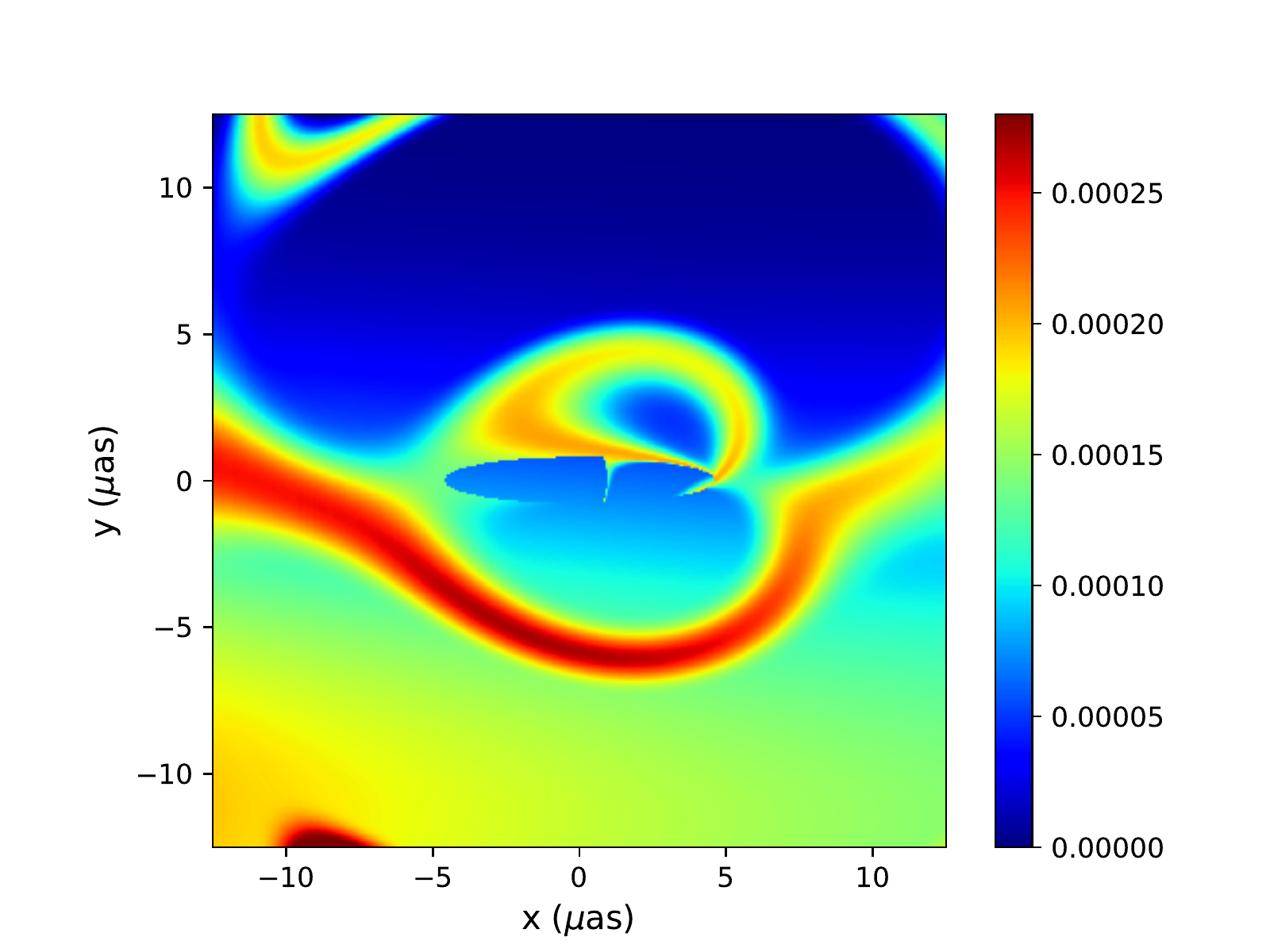}
\end{subfigure}
\end{subsubcaption}

\caption{Images of an accretion torus surrounding a naked rotating wormhole with $a/m=0.9$, $b/m=1$ (upper row) and $a/m=0.9$, $b/m=0.5$ (lower row). The inclination is $\theta=80^{\circ}$ while the field of view is $200\; \mu{\rm as}$ (left column) or  $25\; \mu{\rm as}$ (right column)}
\label{dark_zone}
\end{figure}

Another interesting feature appears when the accretion torus is observed from an inclination angle $\theta$ different from $90^{\circ}$. In this case, the disk $r=0$ located in the equatorial plane becomes visible. It can hardly be seen on the left of Fig. \ref{dark_zone}, but a zoom clearly allows identifying a central dark ellipse\footnote{It is not necessarily an ellipse in the mathematical sense.} on the image on the right. Its contour corresponds to the throat of the wormhole. \\

The blue pixels forming this ellipse-like shape (right panel) represent geodesics coming from $r \rightarrow -\infty$; a similar distorded disk also appears in the case of naked Kerr singularities \cite{HiokiMaeda:2009}. These pixels are not completely black since a part of the torus, located between the throat and the observer, emits some photons directly towards the latter. However there also exists  luminous (green and yellow) pixels inside the dark ellipse. All these illuminated pixels are associated with photons emitted from the torus and travelling through negative values of $r$ back to the observer. The location of this luminous feature inside the dark ellipse highly depends on the value of $b/m$, as is illustrated on the right panel of Fig. \ref{dark_zone}. \\

\begin{figure}[!h]
\begin{center}
\includegraphics[scale=0.5]{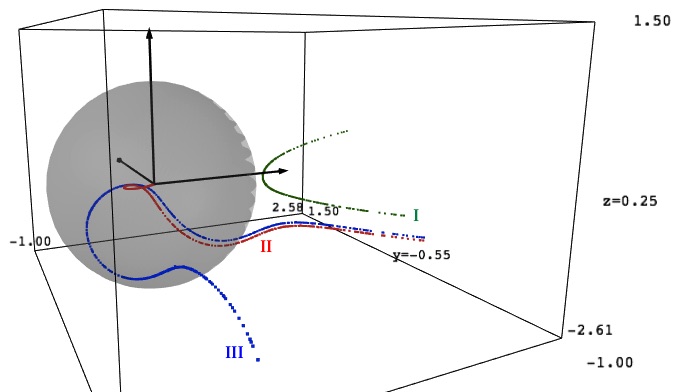}
\caption[]{\label{Geodesics} \footnotesize
The null geodesics associated with three different pixels, starting from the accretion torus for $a/m=0.9$, $b/m=1$, are plotted in a frame with $x=\mbox{e}^r \sin \theta \cos \phi $, $y=\mbox{e}^r \sin \theta \sin \phi$, $z=\mbox{e}^r \cos \theta$. A frame is drawn at the origin, where $r \rightarrow -\infty$. Geodesic I (in green) has a turning point and does not enter the grey sphere of radius $r=0$. Geodesic II (in red) has no turning point, it represents the trajectory of a photon in the central dark ellipse coming from $r \rightarrow -\infty$. Finally, geodesic III (in blue) enters the sphere of radius $r=0$ but has a turning point inside, and goes back to an observer at $r \rightarrow +\infty$.}
 \end{center}
\end{figure}

There exists a sharp contrast between the dark ellipse and this luminous feature, which can be studied in further detail if we consider three different geodesics (Fig. \ref{Geodesics}).

The green geodesic of Fig. \ref{Geodesics} corresponds to a luminous pixel just outside the dark ellipse. This geodesic comes from $r=+\infty$, crosses the torus on its way in, approaches the $r=0$ disk (represented by the grey sphere on Fig. \ref{Geodesics}) without reaching it, and escapes to the observer (crossing a second time the torus on its way out). Similarly, the blue geodesic comes from $r=+\infty$ and crosses the torus on its way in. Contrarily to the green geodesic, it enters the wormhole throat, reaching negative values of $r$. It also reaches a turning point and comes back to $r>0$, also eventually reaching the observer after having crossed the torus a second time on its way out. Finally, the red geodesic originates from $r=-\infty$. It emerges from the throat and crosses the torus only once, on its way out to the observer.

\section{Analytical study of the regular rotating Hayward metric}
\label{section 4}
\subsection{Circular orbits in the equatorial plane}
\subsubsection{Energy and angular momentum of a massive particle}
\leavevmode \par

Let us consider the geodesic motion of a test particle with momentum $p$ and mass $m_0>0$, following a circular orbit in the background of the metric (\ref{metric_improved_Hayward}). This motion occurs in the equatorial plane ($\theta=\pi/2$) due to the axisymmetry of the metric. Along with the property of stationarity, this also implies the existence of two Killing vectors $\xi=\partial_t$ and $\eta=\partial_\phi$. The energy and angular momentum of the particle read:

\begin{equation}
\begin{aligned}
E&=-\xi \cdot p=-p^t (g_{tt}+g_{t\phi}\Omega) \\
L&=\eta \cdot p=p^t(g_{\phi t}+g_{\phi \phi}\Omega),
\end{aligned}
\end{equation}
with $\Omega := \frac{d\phi}{dt}$. \\

The angular velocity $\Omega$ can be found by considering the Euler-Lagrange equation for a free particle whose Lagrangian is

\begin{equation}
\mathscr{L}=\frac{1}{2}g_{\mu \nu}\dot{x}^{\mu}\dot{x}^{\nu},
\end{equation}
with $\dot{x}^\mu:=\frac{dx^\mu}{d \lambda}$, where $\lambda$ is an affine parameter. The Euler-Lagrange equation supplemented by the conditions for a circular orbit $\dot{r}=\ddot{r}=0$ boils down to:

\begin{equation}
g_{tt,r}\dot{t}^2+2g_{t\phi,r}\dot{t}\dot{\phi}+g_{\phi\phi,r}\dot{\phi}^2=0,
\end{equation}
where the coma denotes a derivative with respect to the radial coordinate. \\

The angular velocity of a particle on a circular co- or contra-rotating orbit is then:

\begin{equation} \label{Omega}
\Omega_\pm=\frac{-g_{t\phi,r}\pm \sqrt{g_{t\phi,r}-g_{tt,r}g_{\phi\phi,r}}}{g_{\phi \phi,r}}.
\end{equation}
The specific energy and angular momentum of a massive particle on a circular orbit in a stationary axisymmetric spacetime thus read:

\begin{equation}
\begin{aligned}
\mathcal{E_\pm}& := \frac{E_\pm}{m_0}=-\frac{g_{tt}+g_{t\phi}\Omega_\pm}{\sqrt{-(g_{tt}+2g_{t\phi}\Omega_\pm+g_{\phi\phi}\Omega_\pm^2)}} \\
\mathcal{L_\pm}& := \frac{L_\pm}{m_0}=\frac{g_{\phi t}+g_{\phi\phi}\Omega_\pm}{\sqrt{-(g_{tt}+2g_{t\phi}\Omega_\pm+g_{\phi\phi}\Omega_\pm^2)}}.
\end{aligned}
\end{equation}
In the context of our metric (\ref{metric_improved_Hayward}), we obtain in the equatorial plane ($\theta=\pi/2$):

\begin{equation}
\begin{aligned}
\mathcal{E_\pm}&= \frac{r^3+a^2rM'(r)-M(r)\left(a^2+2r^2\mp2ar\sqrt{A(r)} \right)}{\left(r^3-a^2M(r)+a^2rM'(r)\right)\sqrt{\frac{r^2B_\pm(r)}{\left(r^3-a^2M(r)+a^2rM'(r) \right)^2}}} \\
\mathcal{L_\pm}&=\frac{-(a^3+3ar^2)M(r)+(a^3r+ar^3)M'(r)\pm(a^2r^2+r^4+2a^2rM(r))\sqrt{A(r)}}{\left(r^3-a^2M(r)+a^2rM'(r)\right)\sqrt{\frac{r^2B_\pm(r)}{\left(r^3-a^2M(r)+a^2rM'(r) \right)^2}}},
\end{aligned}
\end{equation}
where
\begin{displaymath}
\begin{aligned}
A(r)&=\frac{M(r)}{r}-M'(r) \\
B_\pm(r)&=-a^2r^2M'^2(r)+r^4-3a^2M(r)^2-3(a^2r+r^3)M(r)+(3a^2r^2+r^4+4a^2rM(r))M'(r) \\
 & \pm 2 \left[(a^3+3ar^2)M(r)-(a^3r+ar^3)M'(r) \right] \sqrt{A(r)}.
\end{aligned}
\end{displaymath}
These expressions differ from Toshmatov et al. \cite{Toshmatov_et_al:2017} (see \ref{s:calculations} for details and a comparison with the results of Bardeen et al. \cite{Bardeen:1972}). \\

\begin{figure}[!h]
\begin{center}
\includegraphics[scale=0.9]{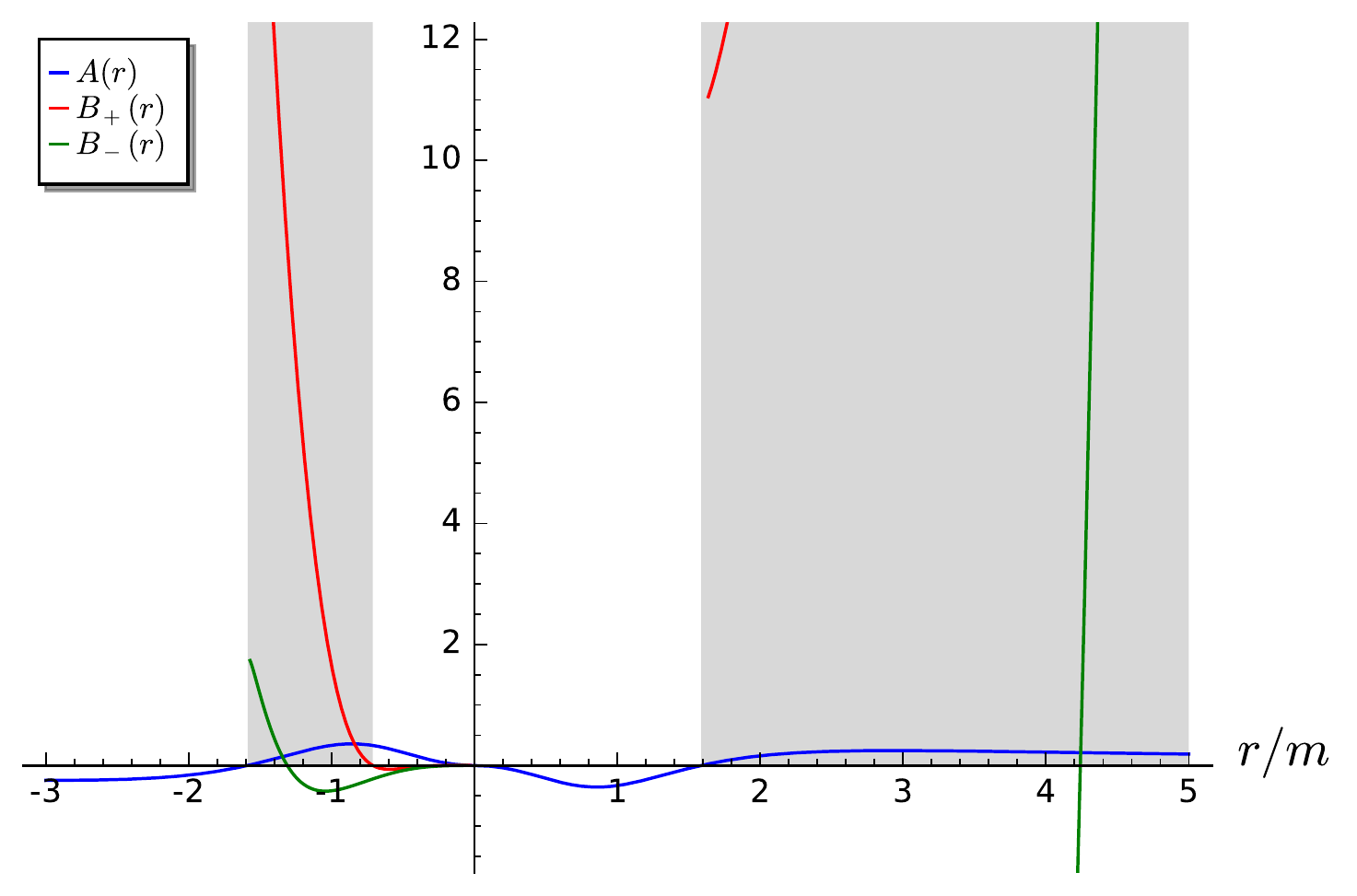}
\caption[]{\label{Circular_orbits_a/m=09_b/m=1} \footnotesize
Plot of $A(r)$, $B_+(r)$ and $B_-(r)$ in the case $a/m=0.9$, $b/m=1$. The shaded regions represent the zones where circular orbits are allowed.}
 \end{center}
\end{figure}

Circular orbits can therefore exist only for $A(r) \geq 0$ and $B_\pm(r)>0$. These three functions are plotted below, for $a/m=0.9$ and $b/m=1$. The regions of allowed co-rotating circular orbits are pictured in grey on Fig.~\ref{Circular_orbits_a/m=09_b/m=1}. The region of positive $r$ goes up to $r \rightarrow +\infty$, while the one of negative values of $r$ exists only near the center. This is coherent, since from $r \rightarrow -\infty$ the metric (\ref{metric_improved_Hayward}) behaves as a Schwarzschild metric with negative mass: the repulsive gravity does not allow circular orbits for large enough negative radii.

\subsubsection{Influence of the spin}
\leavevmode \par

\begin{figure}[!h]
\hspace{-0.9cm}
   \begin{minipage}[c]{.46\linewidth}
      \includegraphics[scale=0.65]{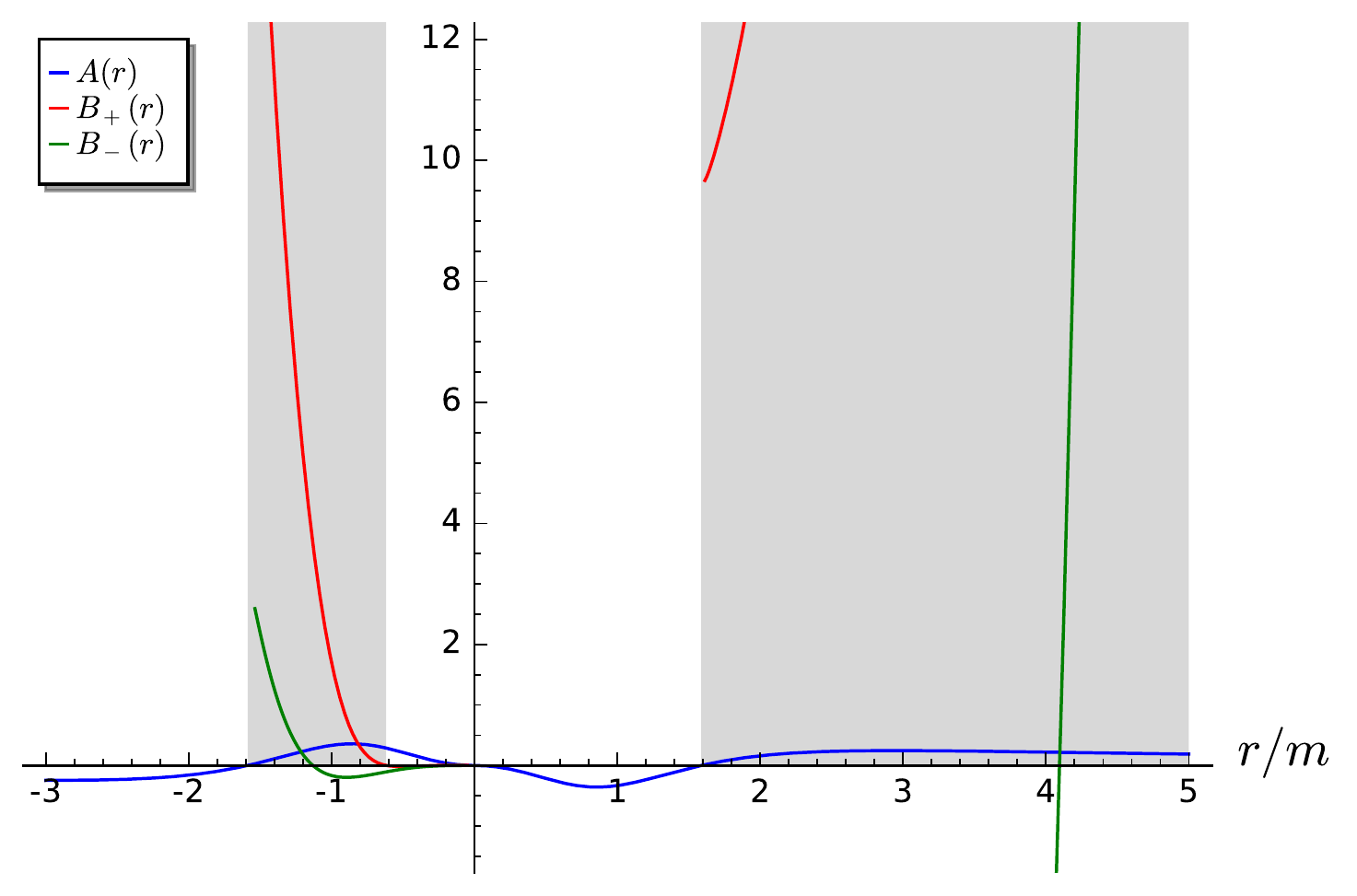}
      \begin{center}
   \footnotesize   a) $a/m=0.8$
      \end{center}
   \end{minipage} \hspace{2cm}
   \begin{minipage}[c]{.46\linewidth}
      \includegraphics[scale=0.65]{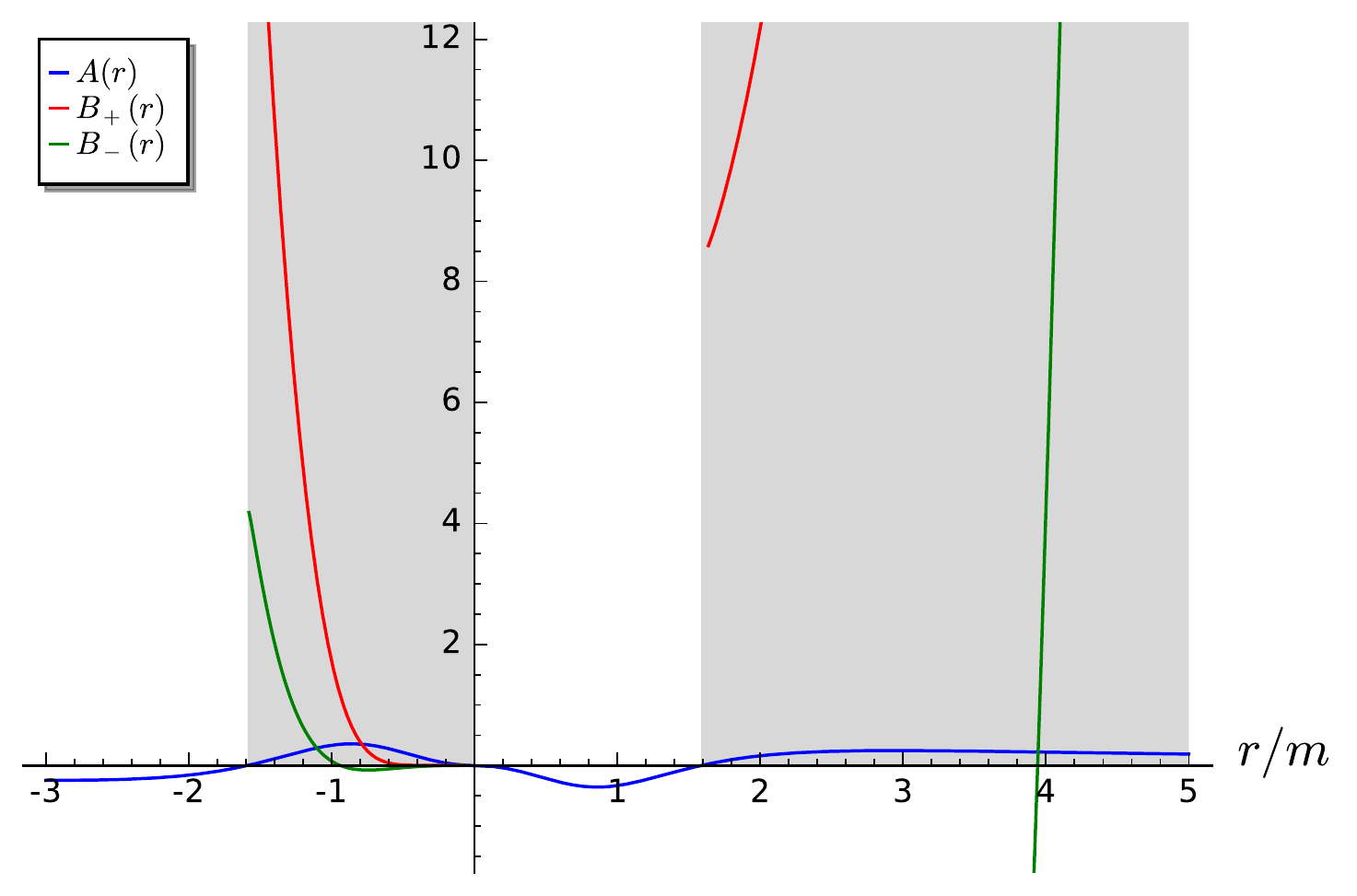}
      \begin{center}
     \footnotesize b) $a/m=0.7$
      \end{center}
   \end{minipage}
   \caption[]{\label{Circular_orbits_a/m=07_b/m=1} \footnotesize
Plot of $A(r)$, $B_+(r)$ and $B_-(r)$ in the case $b/m=1$, for two different values of the spin. The shaded region for the negative values of $r$ gets wider as $a$ decreases.}
\end{figure}

Let us study how the regions of allowed circular orbits are modified when the spin varies.
First of all, it should be noted that $A(r)$ does not depend on the value of the spin. Hence for a fixed $b$, e.g. $b/m=1$ like on Fig.  \ref{Circular_orbits_a/m=09_b/m=1}, the shaded regions will be modified only if $B_\pm(r)$ changes.
As shown on Fig. \ref{Circular_orbits_a/m=07_b/m=1}, decreasing the value of $a$ only widens the zone of circular orbits below $r=0$. It thus does not have any impact on the allowed circular orbits with $r>0$.

\subsubsection{Influence of the parameter $b$}
\leavevmode \par

\begin{figure}[!h]
\hspace{-0.9cm}
   \begin{minipage}[c]{.46\linewidth}
      \includegraphics[scale=0.65]{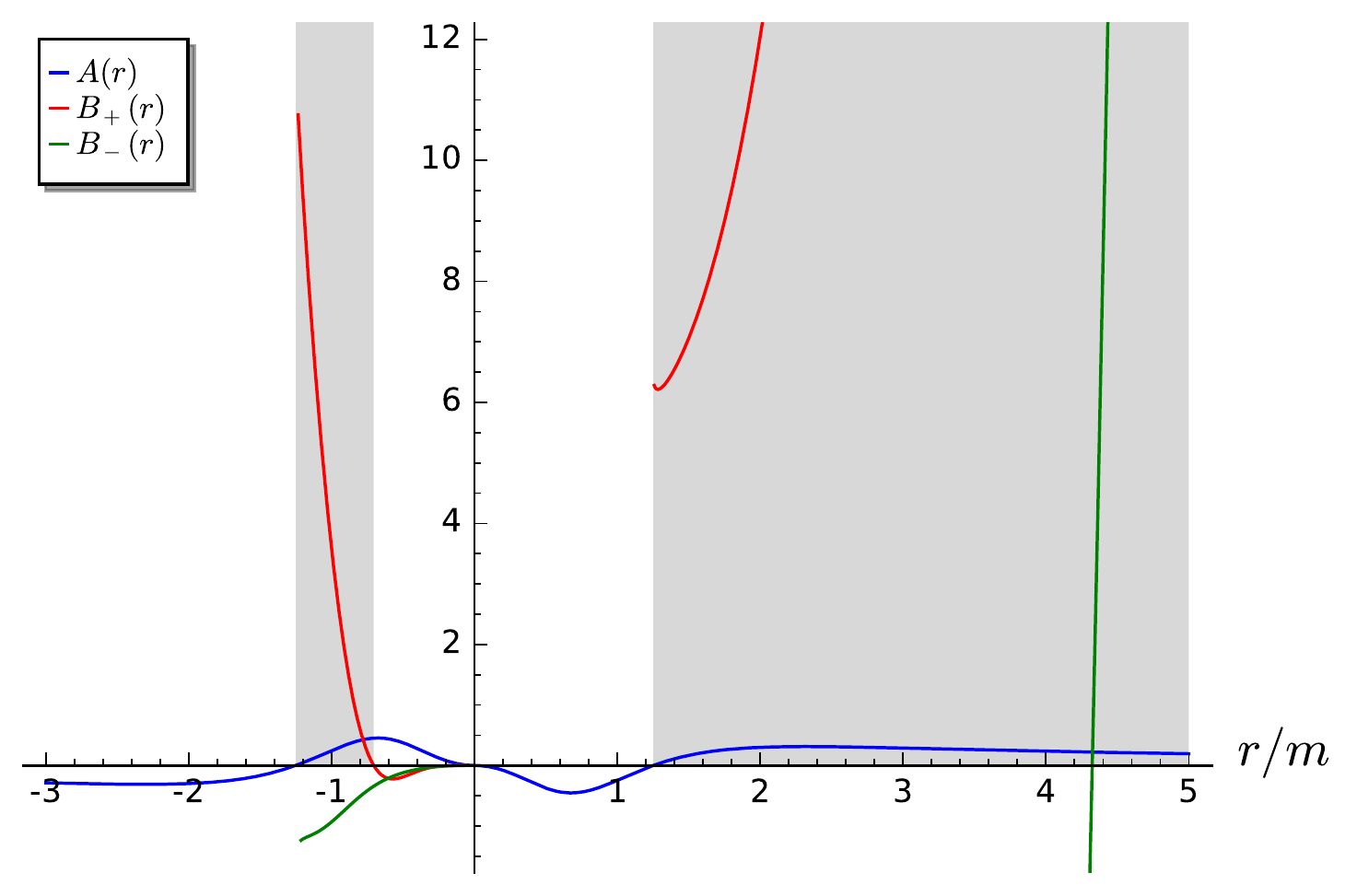}
      \begin{center}
   \footnotesize   a) $b/m=0.7$
      \end{center}
   \end{minipage} \hspace{2cm}
   \begin{minipage}[c]{.46\linewidth}
      \includegraphics[scale=0.65]{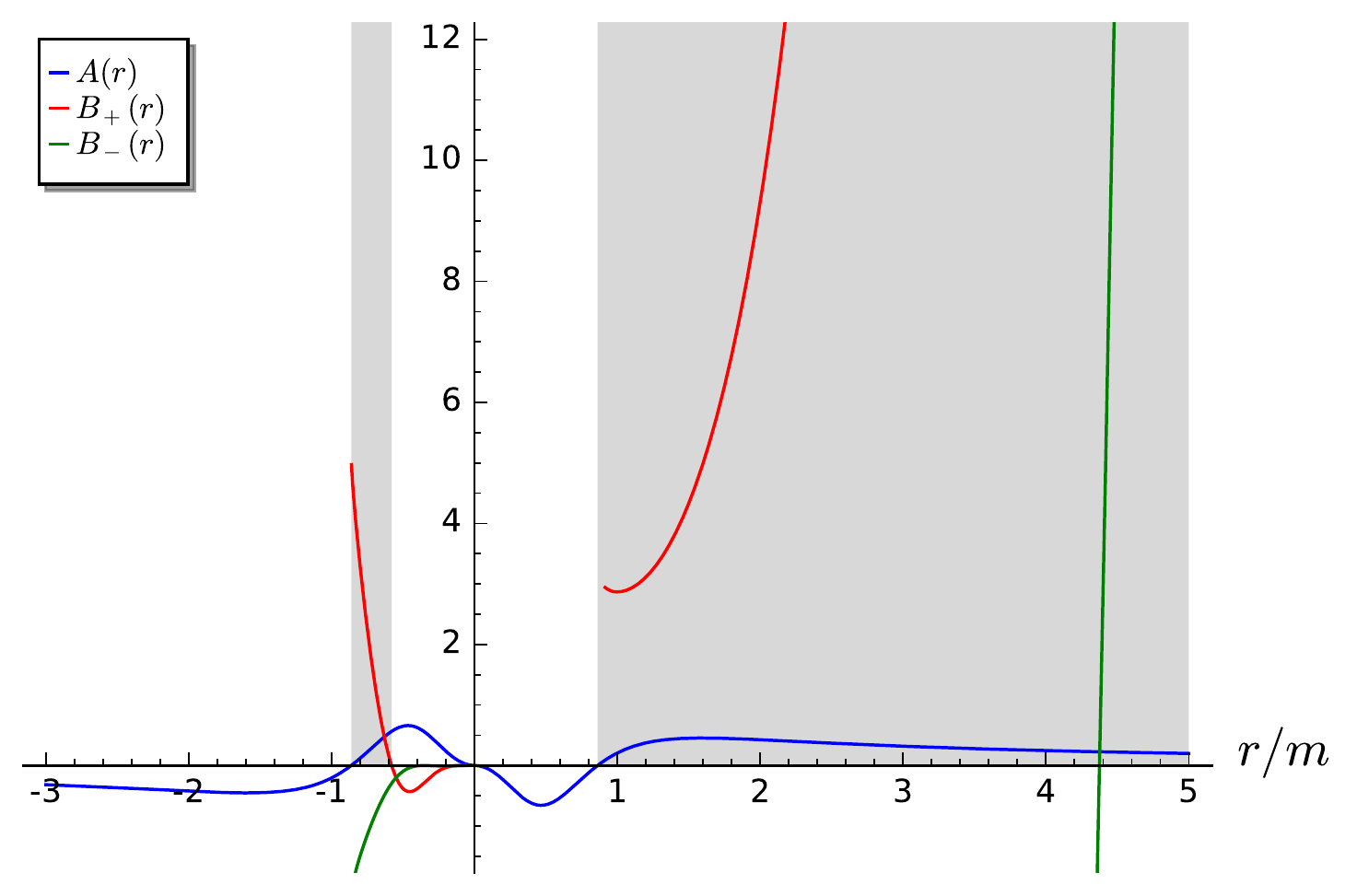}
      \begin{center}
     \footnotesize b) $b/m=0.4$
      \end{center}
   \end{minipage}
   \caption[]{\label{Circular_orbits_a/m=09_b/m=07} \footnotesize
Plot of $A(r)$, $B_+(r)$ and $B_-(r)$ in the case $a/m=0.9$ for two different values of $b$. The shaded region for the negative (resp. positive) values of $r$ gets narrower (resp. wider) as $b$ decreases.}
\end{figure}

Contrarily to the spin, the parameter $b$ has a direct influence on the region of allowed circular orbits of positive radius. Going from $b/m=1$ (Fig. \ref{Circular_orbits_a/m=09_b/m=1}) to $b/m=0.7$ and $b/m=0.4$ (Fig. \ref{Circular_orbits_a/m=09_b/m=07}), at a constant $a/m=0.9$, we observe that circular orbits can occur for smaller and smaller positive values of $r$.

Meanwhile, the region of allowed circular orbits with negative radius shrinks as $b$ decreases. This region even disappears for $b/m=0$, as one can see on Fig. \ref{Circular_orbits_a/m=09_b/m=0} below. In this configuration, two horizons exist and circular orbits occur only for values of the radial coordinate above the radius of the outer horizon. For $b/m=0.2$ (left of Fig. \ref{Circular_orbits_a/m=09_b/m=0}), some circular orbits can also occur below the radius of the inner horizon.

\begin{figure}[!h]
\hspace{-0.9cm}
   \begin{minipage}[c]{.46\linewidth}
      \includegraphics[scale=0.65]{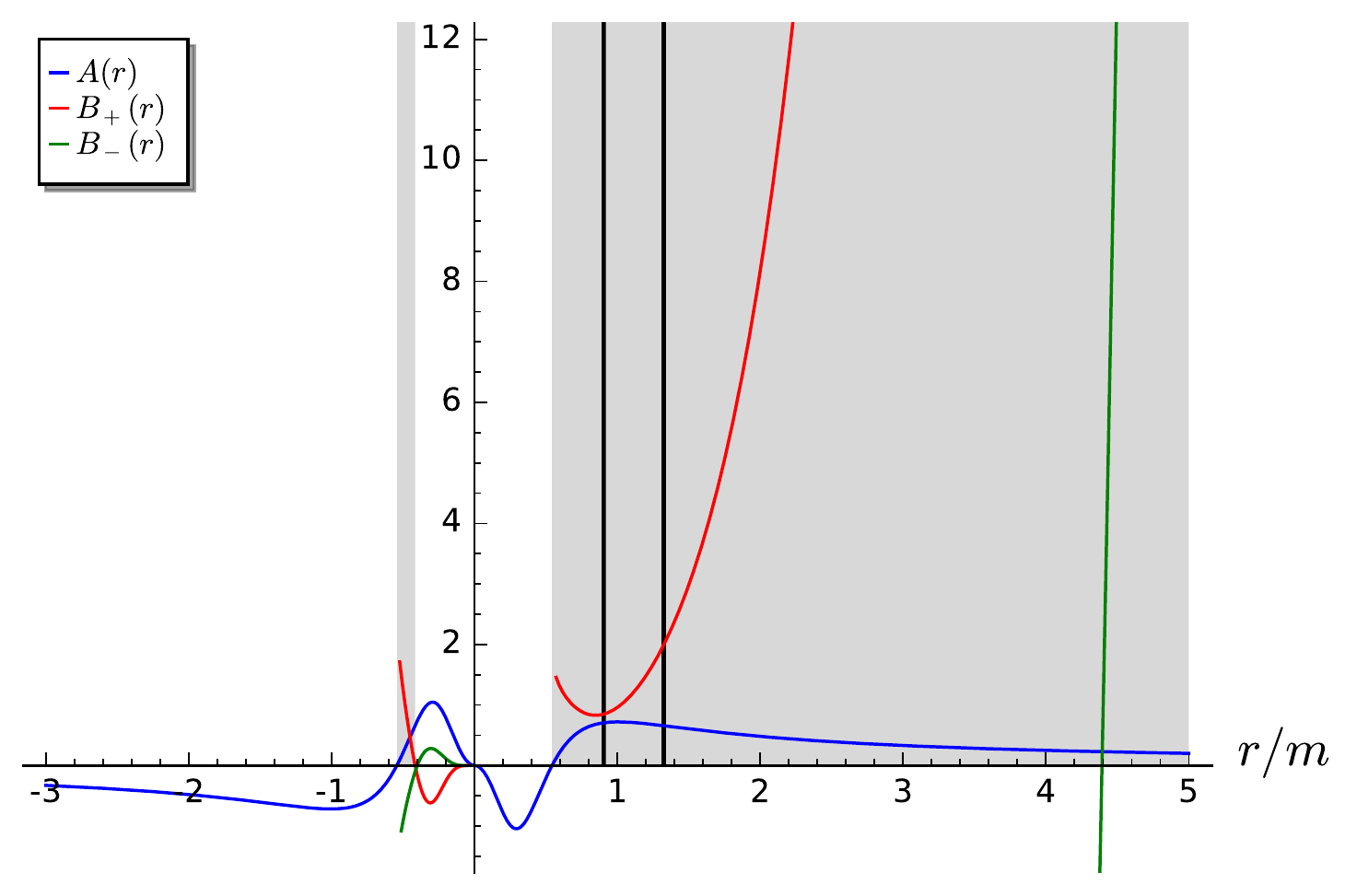}
      \begin{center}
   \footnotesize   a) $b/m=0.2$
      \end{center}
   \end{minipage} \hspace{2cm}
   \begin{minipage}[c]{.46\linewidth}
      \includegraphics[scale=0.65]{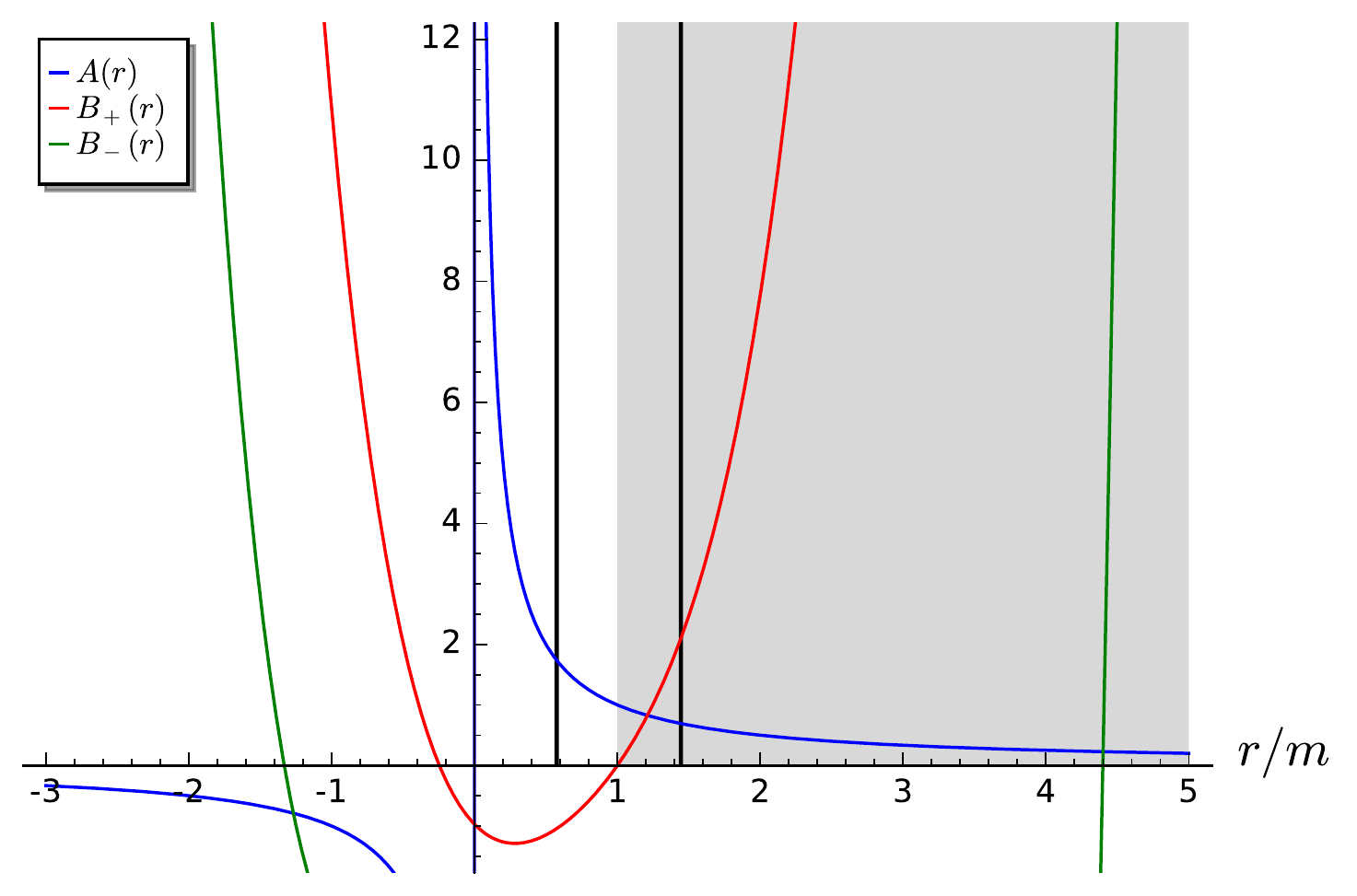}
      \begin{center}
     \footnotesize b) $b/m=0$
      \end{center}
   \end{minipage}
   \caption[]{\label{Circular_orbits_a/m=09_b/m=0} \footnotesize
Plot of $A(r)$, $B_+(r)$ and $B_-(r)$ in the case $a/m=0.9$, in the presence of two horizons (black vertical lines), for two different values of $b$.}
\end{figure}

\subsubsection{Innermost stable circular orbit (ISCO)}
\leavevmode \par

The \emph{innermost stable circular orbit (ISCO)}, which corresponds to the stable circular orbit of smaller $r$, is astrophysically relevant since it provides the highest orbital
frequency possible around the central object. In particular, the ISCO frequency
is involved in various models of quasi-periodic oscillations (QPO) \cite{remillard_x-ray_2006}.

In the Kerr case, it has been shown by Carter \cite{Carter:1968} that the radial geodesic motion
is governed by the following relation:
\begin{equation} \label{def_of_R}
\begin{aligned}
\Sigma \frac{dr}{d\lambda}&=\sqrt{\mathcal{R}} \\
\mathcal{R}&=\left[(r^2+a^2)E-aL \right]^2-\Delta \left[(aE-L)^2+m_0^2r^2+ \mathcal{Q} \right],
\end{aligned}
\end{equation}
where $\lambda$ is an affine parameter, $m_0$ the mass of the particle, and $\mathcal{Q}$, $E$, $L$ are the three integrals of motion (respectively the Carter constant, the energy and the angular momentum of the test particle). The zeros of $\mathcal{R}$ thus represent turning points of the motion of such a test particle in Kerr's spacetime.

Stable circular orbits are defined by the three conditions
\begin{equation}\label{stable_circular_orbits}
\mathcal{R}(r)=0, \quad \frac{d\mathcal{R}(r)}{dr}=0, \quad\mbox{and}\quad \frac{d^2\mathcal{R}(r)}{dr^2} \leq 0 .
\end{equation}

The frequency (\ref{Omega}) of a particle following a circular orbit reads:
\begin{equation} \label{Omega_expl}
\Omega_\pm=\frac{4 \, a b^{2} m^{2} r - a m r^{4} \pm {\left(4 \, b^{4} m^{2} + 4 \, b^{2} m r^{3} + r^{6}\right)} \sqrt{-\frac{4 \, b^{2} m^{2} r^{2} - m r^{5}}{4 \, b^{4} m^{2} + 4 \, b^{2} m r^{3} + r^{6}}}}{r^{7} - {\left(a^{2} - 4 \, b^{2}\right)} m r^{4} + 4 \, {\left(a^{2} b^{2} + b^{4}\right)} m^{2} r}
\end{equation}
The ISCO values of the radius and the orbital frequency (\ref{Omega_expl}), for co-rotating and contra-rotating orbits in the equatorial plane ($\mathcal{Q}=0$), have been computed for different values of $a$ and $b$
(see \ref{s:calculations} for details). The result is shown in Table~\ref{ISCO_a_b}.

\begin{table}[!h]
\begin{center}
\includegraphics[scale=0.184]{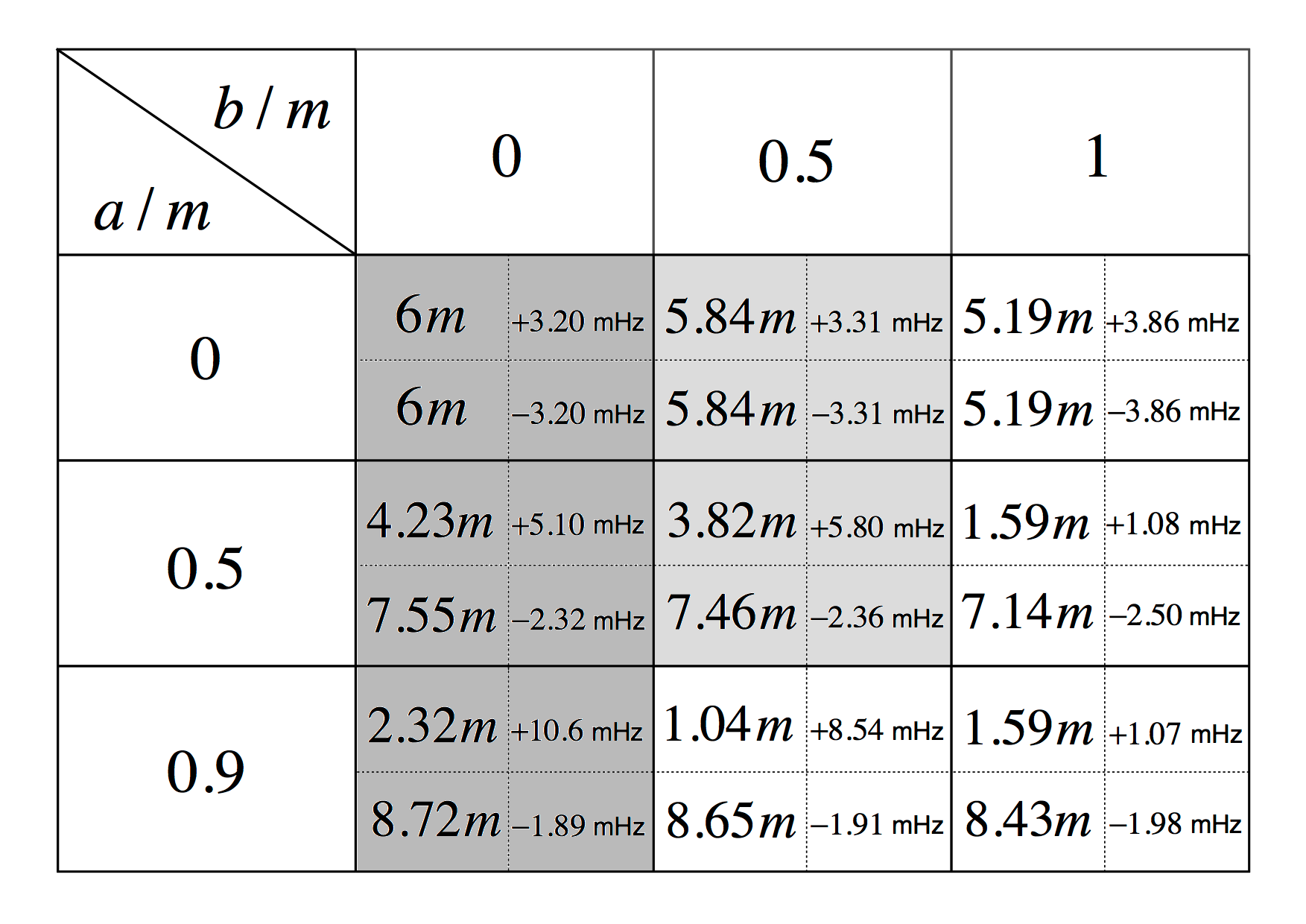}
\caption[]{\label{ISCO_a_b} \footnotesize
Radial coordinate of the ISCO and orbital frequency of a test particle at the ISCO for various values of the parameters $a/m$ and $b/m$, and $m=m_{\mathrm{Sgr\, A}^*}$. Four different results are associated with each combination of $a/m$ and $b/m$: the radius of the ISCO (left) and the frequency $\Omega$ of the co-rotating (resp. contra-rotating) orbit (right) are located on the upper (resp. lower) pannel. The dark grey boxes correspond to the classical Kerr (and Schwarzschild for $a=0$) black hole, the light grey boxes to the regime of rotating regular black hole (zones II and III of Fig. \ref{Causality_Horizon_a_b}), while the others are associated with a naked rotating wormhole (zone IV of Fig. \ref{Causality_Horizon_a_b}).}
 \end{center}
\end{table}

\subsection{Null geodesics}

Let us now focus on the propagation of light rays in order to understand the images that can be seen by an observer on Earth, such as the ones of Fig. \ref{dark_zone}. Due to (\ref{def_of_R}), in which we now take $m_0=0$, the condition for the existence of a photon of energy $E$ with angular momentum $L$ and Carter's constant $\mathcal{Q}$ is
\begin{equation} \label{R>0}
\left[(r^2+a^2)E-aL \right]^2-\left(r^2+a^2-2rM(r) \right) \left[(aE-L)^2+\mathcal{Q} \right] \geq 0 ,
\end{equation}
with $M(r)$ given by Eq.~(\ref{e:M_r_Torres}).

In the case of a Kerr spacetime, $M(r)=m$ and Eq.~(\ref{R>0}) is polynomial
in $r$, of degree 4. It can then be shown that a photon trajectory has at most one radial turning point in the black hole exterior \cite{Frolov&Zelnikov:2015}.
Here, due to the form (\ref{e:M_r_Torres}) of $M(r)$, Eq.~(\ref{R>0}) reduces to a polynomial equation of degree 7. The phenomenology is thus much richer than in Kerr's case. In particular, some photon trajectories can have more than one radial turning point. This is illustrated on Fig. \ref{R_2_turning_points} below. The central shaded region of Fig. (b) is particularly striking: photons with energy $E_0$, angular momentum $L_0$ and a Carter constant $\mathcal{Q}_0$ can oscillate back and forth between two radial turning points, around $r=0$.
\vspace{-3.5cm}
\begin{figure}[!h]
\hspace{-0.9cm}
   \begin{minipage}[c]{.46\linewidth}
      \includegraphics[scale=0.45]{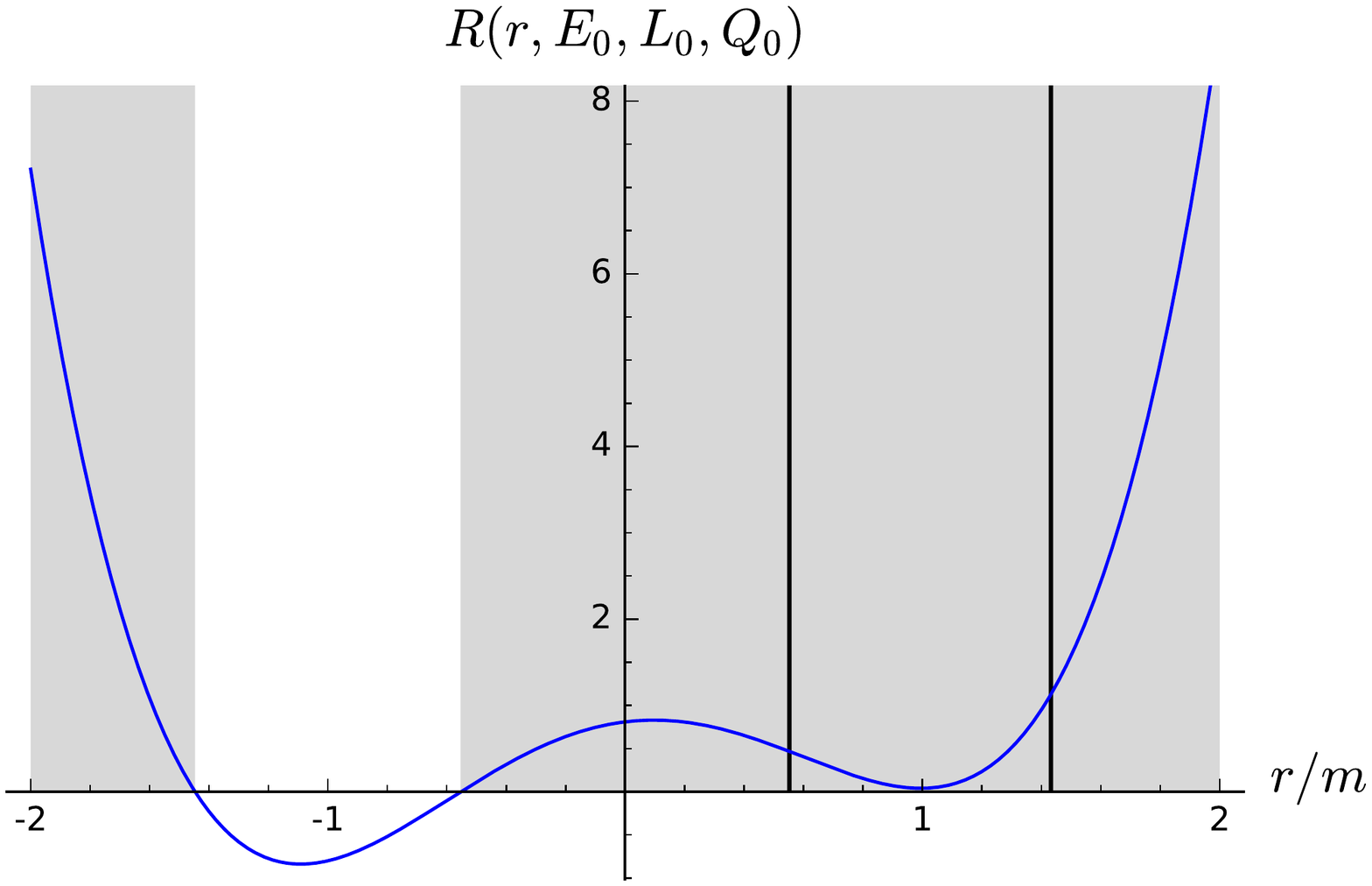}
      \begin{center} \vspace{-4cm}
   \footnotesize   a) $b/m=0$
      \end{center}
   \end{minipage} \hspace{2cm}
   \begin{minipage}[c]{.46\linewidth}
      \includegraphics[scale=0.45]{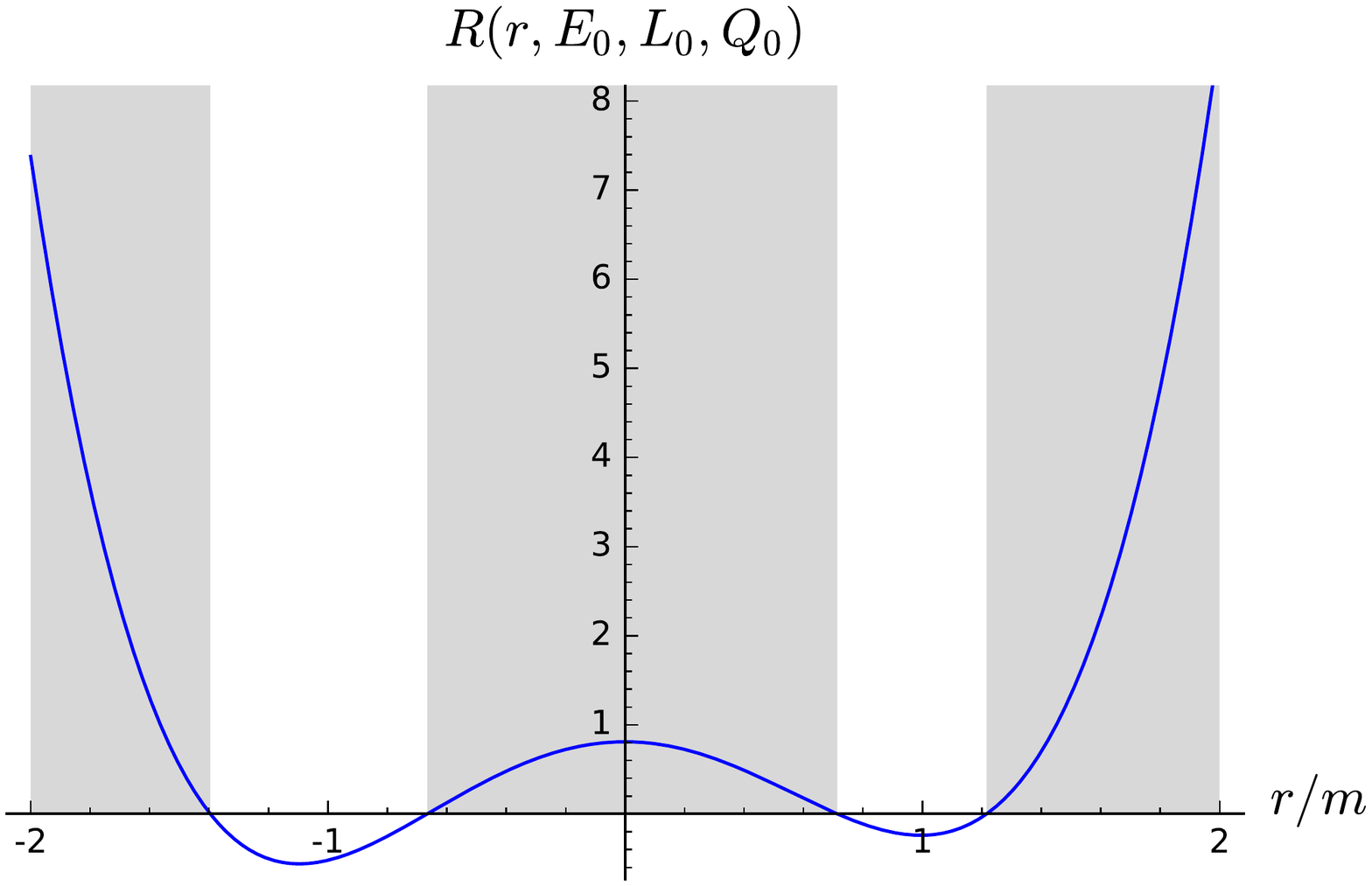}
      \begin{center} \vspace{-4cm}
     \footnotesize b) $b/m=1$
      \end{center}
   \end{minipage}
   \caption[]{\label{R_2_turning_points} \footnotesize
Plot of $\mathcal{R}$ as a function of the radial coordinate $r/m$. The shaded regions represent the allowed regions for a photon with $E_0/m=1$, $L_0/m=2$, $\mathcal{Q}_0/m^2=-1$, in the case of a rotating Hayward black hole with $a/m=0.9$ and $b/m=0$ (a) (the black lines denote the outer and inner horizons) and of a naked rotating wormhole with $b/m=1$ (b).}
\end{figure}

\begin{figure}[!h]
\vspace{-2.5cm}
\hspace{-0.9cm}
   \begin{minipage}[c]{.46\linewidth}
      \includegraphics[scale=0.45]{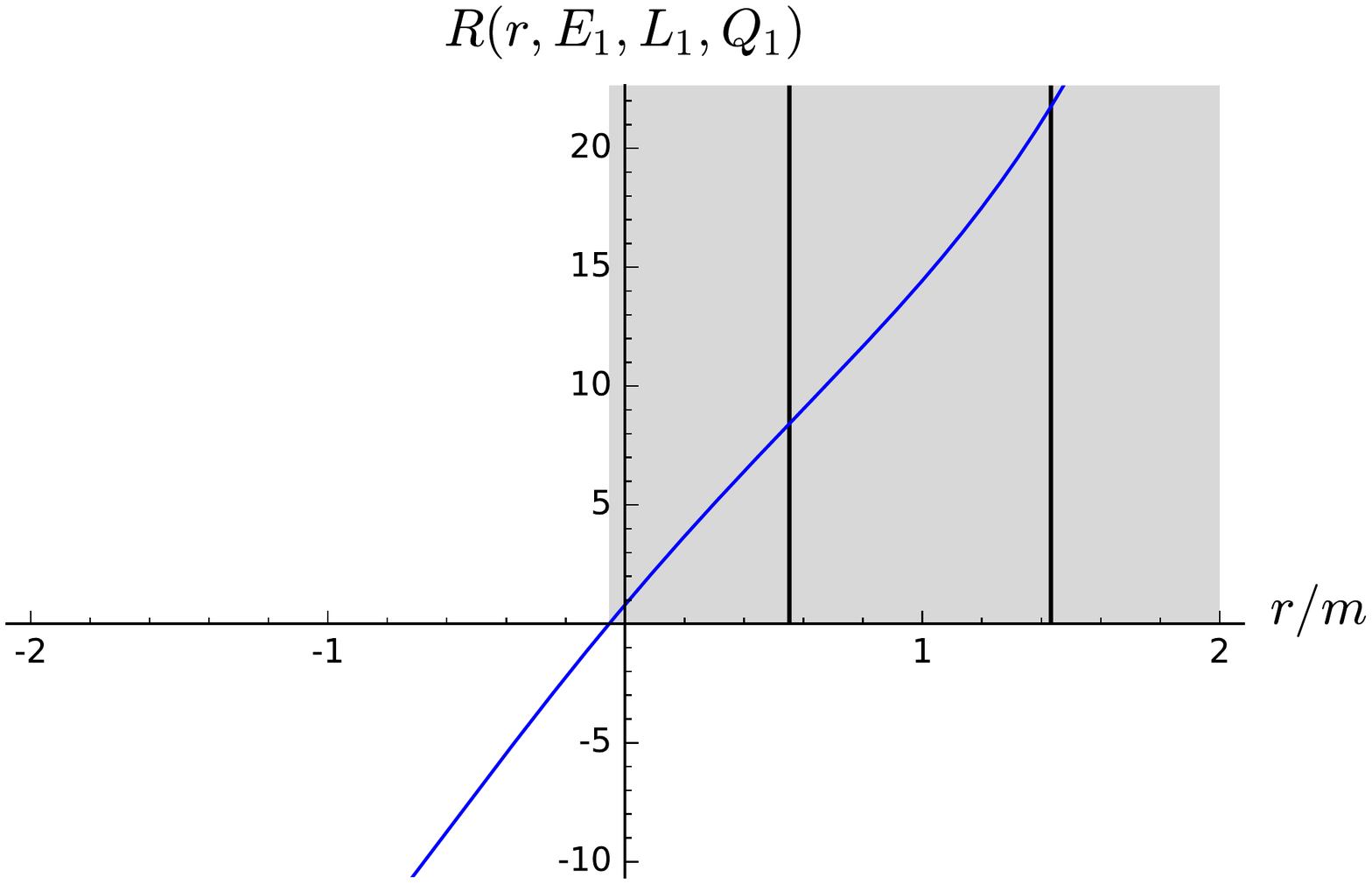}
      \begin{center} \vspace{-4cm}
   \footnotesize   a) $b/m=0$
      \end{center}
   \end{minipage} \hspace{2cm}
   \begin{minipage}[c]{.46\linewidth}
      \includegraphics[scale=0.45]{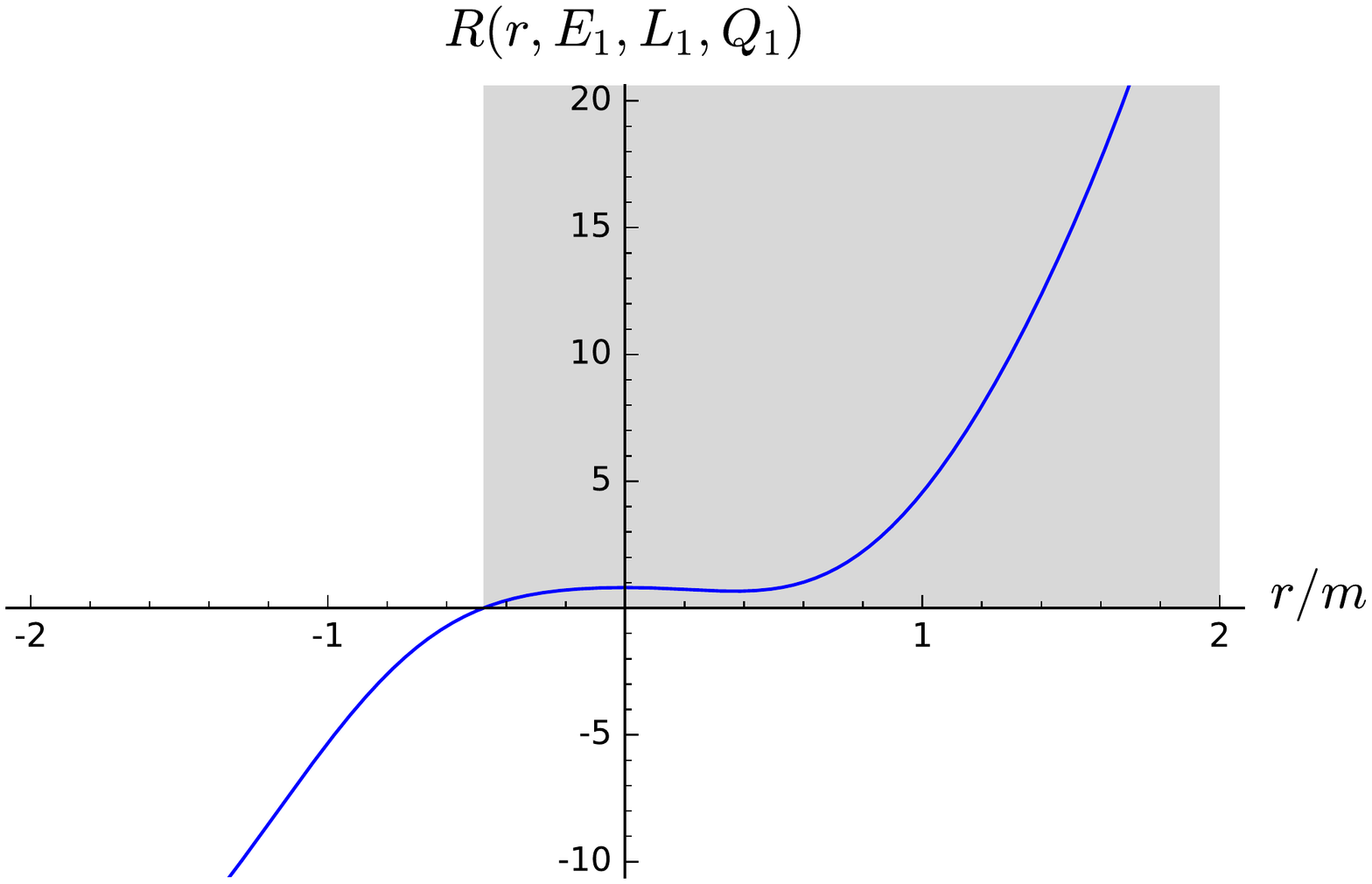}
      \begin{center} \vspace{-4cm}
     \footnotesize b) $b/m=1$
      \end{center}
   \end{minipage}
   \caption[]{\label{R_photons_dark_zone} \footnotesize
Plot of $\mathcal{R}$ as a function of the radial coordinate $r/m$. The shaded regions represent the allowed regions for a photon with $E_1/m=1$, $L_1/m=-2$, $\mathcal{Q}_1/m^2=-1$, in the case of a rotating Hayward black hole with $a/m=0.9$ and $b/m=0$ (a) (the black lines denote the outer and inner horizons) and of a naked rotating wormhole with $b/m=1$ (b).}
\end{figure}

This analysis, using the inequality (\ref{R>0}), also allows us to understand the behaviour of the photons travelling into the region with $r<0$ (Fig. \ref{dark_zone}) before reaching an observer on Earth. Fig. \ref{R_photons_dark_zone} shows the allowed region for a photon with $E_1/m=1$, $L_1/m=-2$, $\mathcal{Q}_1/m^2=-1$ in two different cases: $b/m=0$ (left) and $b/m=1$ (right), while $a/m=0.9$. One can see that in both cases, a photon going from $r>0$ to $r<0$ has a radial turning point and goes back to the region with positive radial coordinate. However, in the case $b/m=0$ where a trapped region is located between the two Killing horizons (black vertical lines), this photon cannot cross the inner horizon and thus reach the observer. When $b/m=1$ no horizon is present, which allows a photon from the accretion torus to travel towards the region $r<0$, reach a turning point and then an observer on Earth.

\newpage

\section{Conclusion}

We have investigated the geometry of a non-singular rotating black hole, both numerically and analytically. To this end we have extended the geodesically incomplete rotating Hayward spacetime \cite{Bambi&Modesto:2013} to the region $r<0$, thereby obtaining a regular rotating Hayward metric. This metric describes a \emph{regular rotating Hayward black hole} in the presence of an event horizon, and a \emph{naked rotating wormhole} otherwise.

The numerical study of the regular rotating Hayward black hole using the ray-tracing code \gy has shown that, at a given spin $a$, the image of an accretion torus around such a black hole possesses a smaller shadow compared to that of the Kerr black hole. This difference is however out of reach of the observations in the foreseeable future.

Some images in the horizonless case (naked rotating wormhole) have also been computed. They display a central faint region with hyper-lensed contours whose shape depends on the value of the parameter $b$. The simulations with \gy allow distinguishing very well these contours from the shadows associated with the standard Kerr case or with the rotating Hayward black hole, as can be seen by comparing Fig. \ref{no_shadow} \& \ref{dark_zone} to Fig. \ref{fig6:a} \& \ref{fig6:b}. Without any good resolution data, we must stress that distinguishing these contours from the lensing ring delineating the shadow of a Kerr black hole could be extremely challenging, as in the case of boson stars \cite{Vincent_et_al:2016}.

Another interesting feature of the naked rotating wormhole occurs when we compute images with
some inclination angle $\theta \neq \pi/2$. A dark ellipse then appears at the centre, corresponding to the image of the wormhole's throat on the observer's sky. This ellipse is mainly dark because there is no source in the region with $r<0$.
However, a luminous feature appears in it, whose shape depends on the value of $b$: it is
produced by photons emitted from the accretion torus, which have crossed the throat and
made some journey in the region $r<0$ before coming back to the observer.

An analytical study of this geometry also has been performed. After giving the expressions for the specific energy and angular momentum of massive particles in co- and contra-rotating orbits in the equatorial plane, which differ from the results of Tomashtov et al. \cite{Toshmatov_et_al:2017}, we computed the radius of the ISCO for various values of the parameters $a$ and $b$. The values of the frequency of the orbits at the ISCO, highly depending on $b$, open up the possibility of distinguishing regular rotating and Kerr black holes  thanks to quasi-periodic oscillations. \\

With the upcoming results of the Event Horizon Telescope, studies of alternatives to Kerr black holes happen to be particularly timely. This  study of a regular rotating black hole is not designed to make a case for the existence of such an object at the center of the Galaxy, especially because it is only an approximate solution of non-linear electrodynamics. But it comes within the scope of previous works on boson stars \cite{Vincent_et_al:2016} or hairy black holes \cite{VincentGourgoulhonHerdeiroRadu:2016} aiming at better understanding the data coming from the EHT. The intriguing similarities observed between boson stars and regular black holes spur to investigate other horizonless geometries in order to make out a common pattern.

\appendix
\section{Calculation details} 

\subsection{SageMath worksheets} \label{s:calculations}

Computation of geometric quantities relative to the metric
(\ref{metric_improved_Hayward})-(\ref{e:M_r_Torres}) have been performed
by means of the free computer algebra system SageMath \cite{SageMath},
thanks to its tensor calculus part (SageManifolds \cite{SageManifolds}).
The corresponding worksheets are available at the following url's:
\begin{itemize}
\item Curvature of the naively extended rotating Hayward metric [Eq.~(\ref{Kretschmann&Ricci})]:\\
{\scriptsize \url{https://cocalc.com/projects/09367c7f-3a39-4079-9d4d-cd59ebdca289/files/Rotating_Hayward_metric_curvature.ipynb}}

\item Curvature of the regular rotating Hayward metric (\ref{metric_improved_Hayward}) extended to the region $r<0$ [Fig.~\ref{R_plot_HT} \& \ref{K_plot_HT}]:\\
{\scriptsize \url{https://cocalc.com/projects/09367c7f-3a39-4079-9d4d-cd59ebdca289/files/rotating_Hayward_metric_ext.ipynb}}

\item Null energy condition in the regular rotating Hayward metric (\ref{metric_improved_Hayward}) extended to the region $r<0$ [Fig.~\ref{NEC_a09_b1}]: \\
{\scriptsize \url{https://cocalc.com/share/09367c7f-3a39-4079-9d4d-cd59ebdca289/Locally_nonrotating_frames_and_NEC.ipynb?viewer=share}}

\item Expressions of the energy, angular momentum and angular velocity of a test particle in the regular rotating Hayward metric (\ref{metric_improved_Hayward}) extended to the region $r<0$; comparison with Toshmatov et al. \cite{Toshmatov_et_al:2017} and Bardeen et al. \cite{Bardeen:1972}: \\
{\scriptsize \url{https://cocalc.com/share/09367c7f-3a39-4079-9d4d-cd59ebdca289/Comparison_of_E_L_and_Omega.ipynb?viewer=share}}

\item Stable circular orbits in the regular rotating Hayward metric (\ref{metric_improved_Hayward}) extended to the region $r<0$ [Table~\ref{ISCO_a_b}]: \\
{\scriptsize \url{https://cocalc.com/projects/09367c7f-3a39-4079-9d4d-cd59ebdca289/files/Stable_circular_orbits.ipynb}}

\end{itemize}

\subsection{Gyoto plugin} \label{Gyoto_plugin}

A Gyoto metric class has been developed inside \gy to obtain all the ray-tracing images displayed in the present paper. It encodes Hayward’s regular rotating metric (10) extended to $r < 0$, and boils down to the metric of a Kerr black hole when $b = 0$. This metric class is part of the standard distribution of Gyoto, freely available at \url{http://gyoto.obspm.fr}.

\section*{References}

\bibliographystyle{unsrt}
\bibliography{biblio}

\end{document}